\documentclass[12pt]{article}

\usepackage{epsfig}

\newcommand{\cz}{\chi_{\scriptscriptstyle Z}}
\newcommand{\GeV}{\mathop{\rm GeV}\nolimits}
\newcommand{\TeV}{\mathop{\rm TeV}\nolimits}
\newcommand{\imag}{\mathop{\rm Im}\nolimits}
\newcommand{\real}{\mathop{\rm Re}\nolimits}
\newcommand{\bbbone}{\hbox{\rm 1\kern-3pt l}}

\begin{document}
\begin{flushright}
MZ-TH/10-47\\
\texttt{1012.4600 [hep-ph]}\\
December 2010\\
\end{flushright}
\vskip 5pt
\begin{center}
\Large{\bf A survey of top quark polarization\\
  at a polarized linear $e^+e^-$ collider}
\end{center}

\begin{center}
\vskip 25pt
{S.~Groote}
\vskip 10pt
{\it \small Loodus- ja Tehnoloogiateaduskond, F\"u\"usika Instituut,\\
  Tartu \"Ulikool, Riia~142, EE--51014 Tartu, Estonia}

\vskip 20pt

{J.G.~K\"orner}
\vskip 10pt
{\it \small Institut f\"ur Physik der Johannes-Gutenberg-Universit\"at,\\
Staudinger Weg 7, D--55099 Mainz, Germany}

\vskip 20pt

{B.~Meli\'c}
\vskip 10pt
{\it \small Rudjer Bo\v skovi\'c Institute, Theoretical Physics Division,\\
Bijeni\v cka c. 54, HR--10000 Zagreb, Croatia}\\
 
\vskip 20pt

{S.~Prelovsek}
\vskip 10pt
{\it \small Physics Department at University of Ljubljana\\
and Jozef Stefan Institute, SI--1000 Ljubljana, Slovenia }

\end{center}

\vskip 20pt

\begin{abstract}
We discuss in detail top quark polarization in above-threshold 
$(t\bar t)$ production at a polarized linear $e^+e^-$ collider. We pay 
particular attention to the minimization and maximization of the polarization
of the top quark by tuning the longitudinal polarization of the $e^+$ and 
$e^-$ beams. The polarization of the top quark is calculated in full
next-to-leading order QCD. We also discuss the beam polarization dependence of
the longitudinal spin--spin correlations of the top and antitop quark spins.
\end{abstract}

\newpage

\section{Introduction}

A future linear $e^+e^-$ collider offers the cleanest conditions for studying
top quark properties, such as the top quark mass, its vector and axial
couplings, and possible magnetic and electric dipole moments. Apart from these
static properties, also the polarization of the top quark can be studied with
great precision. The top decays sufficiently fast so that hadronization
effects do not spoil the polarization which it has at its birth. The large
number of top quark pairs expected to be produced at the ILC, e.g., $50$
$(t\bar t)/{\it hour}$ at $500\GeV$ (based on a luminosity of 
$L=2\times10^{34}cm^{-2}s^{-1}$~\cite{Brau:2007sg,Phinney:2007gp}),
will enable one to precisely determine the top quark polarization from an
angular analysis of its decay products in the dominant decay $t\to X_b+W^+$.
The expected statistical errors in the angular analysis are below the $1\%$
level. Therefore very precise measurements of the angular distributions and
correlations of the decay products of $t$ and $\bar t$ will shed light on the
polarization of the top quarks and on the spin--spin correlations of the top
and antitop quark pairs which are imprinted on the top and antitop quarks by
the $(t\bar t)$-production mechanism. In addition, the measurement of the top
polarization will make it possible to precisely determine the electroweak
Standard Model (SM) parameters or to study a variety of new phenomena beyond
the SM.  

\begin{figure}
\begin{center}
\epsfig{figure=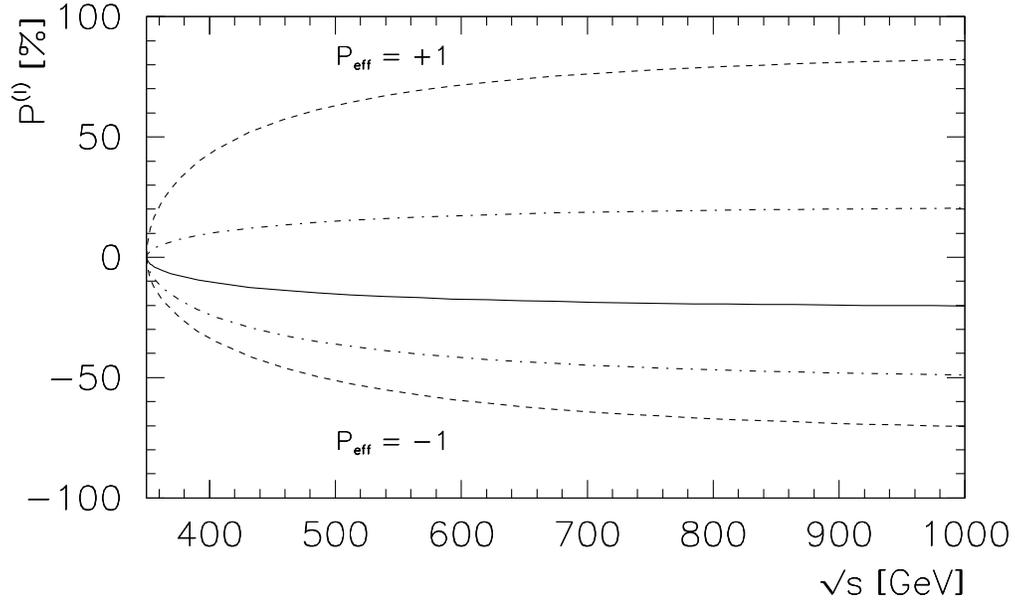, scale=0.8}\\[12pt]\qquad(a)\\[12pt]
\epsfig{figure=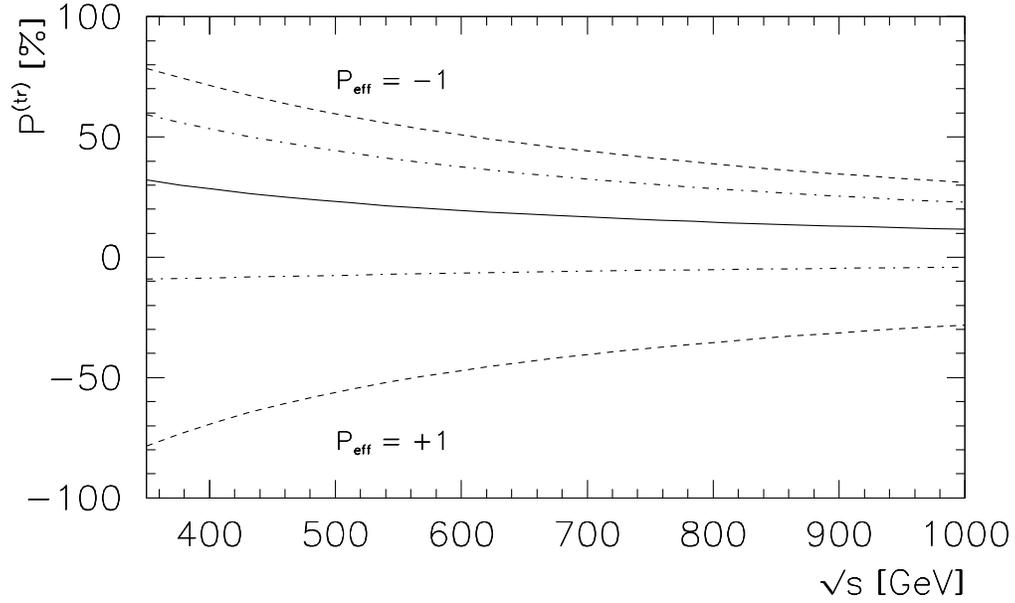, scale=0.8}\\[12pt]\qquad(b)
\end{center}
\caption{Average (a) longitudinal polarization $\langle P^{(\ell)}\rangle$
  and (b) transverse polarization $\langle P^{(tr)}\rangle$ as a function
  of the c.m.\ energy $\sqrt s$, for the values $P_{\rm eff}=-1$ (dashed),
  $P_{\rm eff}=-0.5$ (dash-dotted), $P_{\rm eff}=0$ (solid),
  $P_{\rm eff}=+0.5$ (dash-dotted), and $P_{\rm eff}=+1$ (dashed). Averaging
  is over $\cos\theta$.}
\label{f:fig1}
\end{figure}

It is well known that the top quarks from $e^+e^-$ annihilations are
polarized even for unpolarized $e^+e^-$ beams due to the presence of
parity-violating interactions in the Standard Model (SM). One also knows
from the work of Ref.~\cite{Fischer:1998gsa,rvn00} that the polarization
of the top quark in polarized $e^+e^-$ annihilations can become quite large
when the beam polarization is adequately tuned. This is illustrated in
Fig.~\ref{f:fig1} where we display the energy dependence of the mean
longitudinal and transverse polarization of the top quark in the helicity
system for different values of the effective polarization $P_{\rm eff}$
defined by
\begin{equation}
\label{peff}
P_{\rm eff}=\frac{h_- - h_+}{1 - h_- h_+}\,.
\end{equation}
In Eq.~(\ref{peff}), $h_-$ and $h_+$ are the longitudinal polarization of the
electron and positron beams, respectively. Note that, for unpolarized positron
beams $h_+=0$, one has $P_{\rm eff}=h_-$. For a given value of $h_-$, even
small values of positron polarization of opposite sign will enhance the
effective beam polarization. We shall return to this point in Sec.~2. As
compared to the work of Ref.~\cite{Fischer:1998gsa}, Fig.~\ref{f:fig1} now
includes the $O(\alpha_s)$ radiative corrections. Large single-spin
polarization effects due to beam polarization effects are also implicit in the
work of Parke and Shadmi~\cite{ps96}. Although Ref.~\cite{ps96} is designed for
the analysis of top--antitop quark spin correlations, it is easily adapted to
single-spin polarization effects as also discussed in
Ref.~\cite{Kodaira:1998gt}.

We shall see that the polarization of the top quark is governed by three
parameters: the velocity $v=\sqrt{1-4m^2/s}$, the effective polarization
$P_{\rm eff}$, and the cosine of the scattering angle $\cos\theta$. At the
respective boundaries of the three parameters the description of the
polarization phenomena becomes reasonably simple, in particular, at the Born
term level. The limits $v=0$ (threshold) and $v=1$ (high-energy limit) are
discussed in Sec.~4. In Sec.~5 we discuss the limiting cases
$P_{\rm eff}=\pm1$.

The two respective limiting cases $v=0,1$ and $P_{\rm eff}=\pm1$ contain in a
nutshell much of the information that we want to discuss in the remaining part
of the paper for intermediate values of these parameters. Many of the
qualitative features of our results can be understood from extrapolations
away from the two respective limits.

We shall also address the question of how to maximize and minimize the
polarization of the top quark by tuning the beam polarization. Whereas a
maximum polarization is optimal for the experimental determination of
polarization effects, it is often desirable to gauge the quality of a
polarization measurement against the corresponding unpolarized decay analysis.
For some measurements it may even be advantageous to eliminate polarization
effects altogether. 

Of course, in the tuning process one has to bear in mind to keep the
production rate at an acceptable level. This problem is not unrelated to the
one of the original motivations of including beam polarization in linear
colliders, namely, the gain in rate through beam polarization effects. We
shall also address this question.

Our paper is structured as follows. In Sec.~2 we present the spin formalism
of polarized beam production of top--antitop quark pairs including the polar
angle dependence of the various spin components and longitudinal beam
polarization effects. We present Born term and loop formulas for the relevant
structure functions and collect general expressions necessary for the
numerators and the denominator of the polarization observables. Section~3
contains numerical next-to-leading (NLO) results on the angle-integrated rate
and on polar angle distributions of the rate including beam polarization
effects. We also provide numerical results on the left--right polarization
asymmetry $A_{LR}$. In Sec.~4 we discuss the limiting cases $v=0$ and $v=1$ at
the Born term level. In Sec.~5 we describe the simplifications that occur for
maximal effective beam polarizations $P_{\rm eff}=\mp1$ which correspond to
the $(e^-_{L/R},e^+_{R/L})$ beam configurations. In Sec.~6 we discuss beam
polarizations effects on the three components of the top quark polarization
vector. Section~7 contains a discussion of the magnitude and the orientation of
the polarization vector of the top quark. In Sec.~8 we present numerical
NLO results on beam polarization effects on longitudinal spin--spin
correlations of the top and antitop quark. Finally, Sec.~9 contains a summary
of our results and our conclusions. In an Appendix we list the electroweak
coupling coefficients used in this paper and relate them to the chiral
electroweak coupling coefficients used e.g.\ in Ref.~\cite{ps96}.

Many of the quantitative arguments presented in this paper are based on Born
term level results for which we give explicit alpha-numerical expressions for
$\sqrt s=500\GeV$. We emphasize, though, that all numerical results presented
in the plots include the full $O(\alpha_s)$ radiative corrections where we
have integrated over the full gluon phase space. By comparing the graphical
NLO results with the numerical LO results, one can assess the size of the
$O(\alpha_s)$ radiative corrections, at least for the representative energy of
$\sqrt s = 500\GeV$. In general, the $O(\alpha_s)$ corrections to polarization
observables are small (up to several percent) but can become much larger in
some areas of phase space. A case in point is the longitudinal polarization of
the bottom quark produced on the $Z^0$ at the backward point which obtains a
$25\%$ $O(\alpha_s)$ correction when $P_{\rm eff}=+1$~\cite{rvn00}. As we shall
see later on, the  $O(\alpha_s)$ corrections to $(t\bar t)$-production can
amount up to $12\%$ (see Sec.~7). In addition, there are polarization
observables that are zero at the Born term level and become populated only at
$O(\alpha_s)$. Among these are the normal component of the polarization (see
Sec.~6) and the longitudinal polarization produced from a longitudinal
intermediate vector boson (see Sec.~2). 

\section{Spin formalism of polarized beam production}
The production of top quark pairs at a linear $e^+e^-$-collider proceeds via
$\gamma$- and $Z$-exchange,
\begin{equation}
e^-e^+ \stackrel{\gamma\,,Z}\rightarrow t\bar t\,.
\end{equation}
At the center of mass energies which are being envisaged at the ILC
($s=(p_{e^-}+p_{e^+})^2$), $\sqrt s\sim 2m_t\div 1000\GeV$, top quark pairs
will be produced with nonrelativistic velocities in the threshold region
($v\to 0$) up to relativistic velocities of $v=0.937$ at the highest energy
$\sqrt s=1000\GeV$ $(v=\sqrt{1-4m_t^2/s}$).\footnote{In the first stage of
the ILC, one will reach energies up to $500\GeV$ with an optional second stage
upgrade to $1000\GeV$~\cite{Brau:2007sg,Phinney:2007gp}. For the multi-TeV
collider CLIC one foresees energies up to $3\TeV$~\cite{Assmann:2000hg}.}
This enables the study of the complete production phenomena with different
polarization and correlation effects that reach from the nonrelativistic to
the relativistic domain. For unpolarized beams the total rate is dominated by
the diagonal ($\gamma$ -- $\gamma$) and the ($Z$ -- $Z$) rates which
contribute at the same order of magnitude. The ($\gamma$ -- $Z$) interference
contribution to the total rate is suppressed due to the smallness of the
vector $(Z\,e^+e^-)$ coupling $v_l$ ($v_l=-1+4\sin^2\theta_W$). The
($\gamma$ -- $Z$) interference contribution can, however, become quite sizable
for polarized beams, for the polar angle dependent rates and for top quark
polarization effects.

We mention that, at threshold, there will be the opportunity for very precise
measurements of the top quark mass and width, as well as of the strong
coupling $\alpha_s$. In this region, perturbative QCD is no longer applicable.
One has to solve the Schr\"odinger equation for the relevant Green functions
in a nonrelativistic approximation for a Coulombic potential, i.e.\ the
nonrelativistic QCD (NRQCD) method, described first in Ref.~\cite{FK} and
later applied to the calculation of various different quantities at threshold
(see for example the discussion in Ref.~\cite{Harlander:1996vg,AJ} and
references therein). In this paper we shall discuss top--antitop production
well above threshold where perturbation theory can be safely applied. For our
purposes we take the perturbative regime to start approximately $10\GeV$ above
threshold. Throughout this paper we shall take the top quark mass to have a
nominal value of $175\GeV$. Therefore, we shall consider c.m.\ beam energies
starting from $\sqrt s=360\GeV$.

We are going to discuss the most general case of the polarization of the top
quark with arbitrary longitudinal polarizations of the $e^-$- and $e^+$-beams.
The rate depends on the set of four parameters
$\{h_-\in[-1,1],\,h_+\in[-1,1],\,v\in[0,1],\,\cos\theta\in[-1,1]\}$ or,
equivalently, on the set $\{K_G\in[0,2],\,P_{\rm eff}\in[-1,1],\,v\in[0,1],
\,\cos\theta\in[-1,1]\}$, where we shall call $K_G=1-h_-h_+$ the gain factor.
We have indicated the range of the parameter values in square brackets. In
contrast to the rate, the polarization of the top quark depends only on the
set of the three parameters $\{P_{\rm eff},v,\cos\theta\}$. When discussing
our predictions we shall attempt to explore the whole four- and
three-dimensional parameter space for the rate and polarization, respectively.
We mention that the beam polarizations envisaged at the ILC are $h_-=\pm 90\%$
for electrons and $h_+=\pm 80\%$ for positrons~\cite{Alexander:2009nb}.

We will see that beam polarizations significantly influence the polarization
phenomena of a top quark. In addition, adequately tuned beam polarization can
enhance the top--antitop quark signal and suppress other background processes
such as $W$-pair production (see discussion in
Ref.~\cite{MoortgatPick:2005cw}).

In what follows, we concentrate on the polarization of the top quark, i.e.\ we
sum over the polarization of the antitop quark. The polarization of the antitop
quark can be obtained from the corresponding polarization components of the
top quark using CP invariance as will be discussed in the summary section. Even
more structure is revealed when one considers joint top-antitop polarization.
In  order to reveal this structure, one must perform a joint analysis of the
decay products of the top and antitop quark. $(t\bar t)$ spin--spin
correlations will be briefly discussed in Sec.~8 at the end of the paper.

\begin{figure}
\begin{center}
\includegraphics[width=0.9\textwidth]{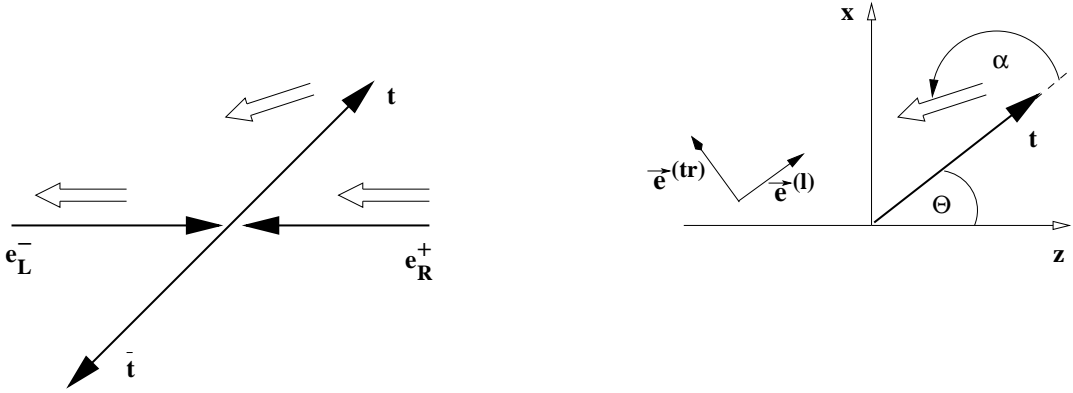}
\end{center}
\caption{A generic configuration for top pair production and top polarization
at a polarized $e^+ e^-$ collider with a $(e^-_{L},e^+_{R})$ polarization. The
positive $z$ axis points into the direction of the electron momentum. The
angle $\alpha$ is the polar angle of the top quark polarization relative to
the top quark momentum measured anticlockwise from the direction of the top
quark.\label{f:fig2}}
\end{figure}

The general expression of the cross section for $(t\bar t)$ production in
$e^+e^-$ collisions is given by\footnote{The spin kinematics of
$e^+e^-$ collisions has been formulated in a number of papers. These include
the unpublished DESY report~\cite{ks81} of which the portions relevant to this
paper have been summarized in Ref.~\cite{gkt97}. Other papers on the subject
are Refs.~\cite{rvn00,MoortgatPick:2005cw,re81,hi86,ha94,krz86}.}
\begin{equation}
\label{sigmam}
d\sigma^{(m)}=2\pi\frac{e^4}{s^2}\sum_{i,j=1}^4
  g_{ij}L^{i\,\mu\nu}H_{\mu\nu}^{j(m)}dPS\,.
\end{equation}
$L_{\mu\nu}^i$ is the lepton tensor, $H_{\mu\nu}^j$ is the hadron tensor
encoding the hadronic production dynamics, $dPS$ is the phase space factor,
and the $g_{ij}$ are the elements of the electroweak coupling matrix which are
defined in the Appendix. The sum runs over the four independent configurations
of products of the vector and axial vector currents, i.e.\ $i,j=1,2$ for
$(VV\pm AA)/2$, \,$i,j=3$ for $i(V\!A-AV)/2$, and $i,j=4$ for $(V\!A+AV)/2$ for
the product of lepton and quark currents, and $m$ denotes one of the possible
polarization configurations of the top quark: longitudinal ($m=\ell$),
transverse ($m=tr$) in the beam scattering plane and normal ($m=n$) to the
beam scattering plane. Our choice of the three orthonormal spin directions
$(\vec e^{\,(tr)},\vec e^{\,(n)},\vec e^{\,(\ell)})$ are given by
\begin{equation}\label{basistop}
\vec e^{\,(tr)}=\frac{(\vec p_{e^-}\times\vec p_t)\times\vec p_t}
{|(\vec p_{e^-}\times\vec p_t)\times\vec p_t|},\qquad
\vec e^{\,(n)}=\frac{\vec p_{e^-}\times\vec p_t}
{|\vec p_{e^-}\times\vec p_t|},\qquad 
\vec e^{\,(\ell)}=\frac{\vec p_t}{|\vec p_t|}\,.
\end{equation}
In Fig.~\ref{f:fig2} we have drawn the directions of $\vec e^{\,(tr)}$
and $\vec e^{\,(\ell)}$ for a generic top quark direction; the vector
$\vec e^{\,(n)}$ shows out of the plane. For the unpolarized top quark case
the superscript $(m)$ is dropped in Eq.~(\ref{sigmam}). The explicit
definitions for all the above quantities together with explicit analytical
expressions for the radiative corrections can be found in
Refs.~\cite{gkt97,kpt94,gkt96,gk96} (see also Ref.~\cite{rvn00}).

We proceed with the discussion in the helicity basis, i.e.\ we take the
direction of the top quark to define the $z$ direction of the hadronic system.
For unpolarized beams the angular decomposition of the differential polarized
cross section can be written as
\begin{eqnarray}
\label{diffcross}
\frac{d\sigma^{(m)}}{d\cos\theta}&=&\frac38(1+\cos^2\theta)\sigma_U^{(\ell)}
  +\frac34\sin^2\theta\sigma_L^{(\ell)}+\frac34\cos\theta\sigma_F^{(\ell)}
  \nonumber\\&&\strut
  -\frac3{\sqrt2}\sin\theta\cos\theta\sigma_I^{(tr,n)}
  -\frac3{\sqrt2}\sin\theta\sigma_A^{(tr,n)}\,,
\end{eqnarray}
where, at NLO of QCD,
\begin{equation}
\label{uli}
\sigma_a^{(m)}=\frac{\pi\alpha^2v}{3s^2}\sum_{j=1}^4
  g_{1j}(H_a^{j(m)}({\it Born\/})+H_a^{j(m)}(\alpha_s))
  \qquad\qquad\mbox{for $a=U,L,I$}
\end{equation}
and
\begin{equation}
\label{fa}
\sigma_a^{(m)}=\frac{\pi\alpha^2v}{3s^2}\sum_{j=1}^4
  g_{4j}(H_a^{4(m)}({\it Born\/})+H_a^{4(m)}(\alpha_s))
  \qquad\qquad\mbox{for $a=F,A$}\,.
\end{equation}
In Eq.~(\ref{diffcross}) we have rewritten the covariant representation 
(\ref{sigmam}) in terms of helicity structure functions $\sigma_a^{(m)}$. The
angle $\theta$ is the polar angle between the momentum of the top quark and
the electron momentum (see Fig.~\ref{f:fig2}). For example, in the purely
electromagnetic case $e^+e^-\to\gamma^*\to q\bar q$ one obtains the LO formula
\begin{equation}\label{pdgform}
\frac{d\sigma}{\cos\theta}=2\pi N_cQ_f^2v\frac{\alpha^2}{4s}
\Big(1+\cos^2\theta+(1-v^2)\sin^2\theta\Big)
\end{equation}
using the LO born term expressions listed later in Eq.~(\ref{HUL1}). The
distribution~(\ref{pdgform}) agrees with Eq.~(41.2) in the PDG booklet.
We mention that our $O(\alpha_s)$ corrections agree with those in
Ref.~\cite{rvn00} after correcting a sign mistake in the normal polarization
(see Erratum in Ref.~\cite{gk96}). 

Above the top quark threshold, one is sufficiently far away from the $Z$-boson
pole to neglect the imaginary part of the $Z$ boson pole propagator. This can
be appreciated from the Breit-Wigner line shape of the $Z$ propagator,
{\it viz.}
\begin{equation}
\chi_Z=\frac1{s-M_Z^2+iM_Z\Gamma_Z}=\frac1{s-M_Z^2}
\Big(1-i\frac{M_Z\Gamma_Z}{s-M_Z^2}\Big)\Big/
\Big(1+\frac{M_Z^2\Gamma_Z^2}{(s-M_Z^2)^2}\Big).
\end{equation}
The factor $M_Z\Gamma_Z/(s-M_Z^2)$ determines the ratio of the imaginary and
real parts of the $Z$ propagator $\imag\chi_Z/\real\chi_Z$. It is already
quite small at threshold ($\sim 0.002$) and falls off with $s^{-1}$. Dropping
the imaginary part contribution of the $Z$ propagator implies that we neglect
contributions proportional to $g_{13}$ in Eq.~(\ref{uli}) and $g_{43}$ in
Eq.~(\ref{fa}). We shall also neglect the width dependence in the real part of
the $Z$ propagator because it is negligibly small.

The nonvanishing unpolarized Born term contributions $H_a^j(Born)$ read
\begin{eqnarray}\label{HUL1}
H_U^1({\it Born\/})=2N_cs(1+v^2),&&
H_L^1({\it Born\/})=N_cs(1-v^2)\ =\ H_L^2({\it Born\/}),\nonumber\\[3pt]
H_U^2({\it Born\/})=2N_cs(1-v^2),&&
H_F^4({\it Born\/})=4N_csv.
\end{eqnarray}
One has $(1-v^2)=4m_t^2/s$ showing that the longitudinal rate $H_L$ falls off
with a $s^{-1}$ power behaviour relative to the transverse rates $H_{U,F}$.
The longitudinally polarized contributions read
\begin{eqnarray}
\label{ell}
H_U^{4(\ell)}({\it Born\/})=4N_csv,&&
H_F^{1(\ell)}({\it Born\/})=2N_cs(1+v^2),\nonumber\\[3pt]
H_L^{4(\ell)}({\it Born\/})=0,&&
H_F^{2(\ell)}({\it Born\/})=2N_cs(1-v^2).
\end{eqnarray}

Note that one has the Born term relations
\begin{eqnarray}
\label{identities}
H_U^{4(\ell)}({\it Born\/})&=&H_F^4({\it Born\/})\,,\nonumber \\
H_F^{1,2(\ell)}({\it Born\/})&=&H_U^{1,2}({\it Born\/})\,,
\end{eqnarray}
which are due to angular momentum conversation in the back-to-back
configuration at the Born term level. It is quite clear that
Eqs.~(\ref{identities}) no longer hold true in general at $O(\alpha_s)$ since
quark and antiquark are no longer back-to-back in general due to additional
gluon emission. The relations~(\ref{identities}) will be useful in our
subsequent discussion of the longitudinal polarization at the forward and
backward points. Notable also is the relation $H_L^{4(\ell)}({\it Born\/})=0$
in Eq.~(\ref{ell}) which is again related to the LO back-to-back configuration.
The radiative corrections to the corresponding polarization component
$P_{L}^{(\ell)}$ have been studied in Ref.~\cite{gkt96} and have been found to
be small of $O(0.1\%)$ when averaged over gluon phase space. For small top
quark energies $P_{L}^{(\ell)}$ can become as large as $3\%$ at
$\sqrt s=500\GeV$.

For the transverse polarization components, one has~\cite{gk96}
\begin{equation}
H_I^{4(tr)}({\it Born\/})=2N_csv\frac{m_t}{\sqrt{2s}},\quad
H_A^{1(tr)}({\it Born\/})=2N_cs\frac{m_t}{\sqrt{2s}}
  =H_A^{2(tr)}({\it Born\/})\,.
\end{equation}
The only nonnegligible contribution to the normal polarization component 
$P^{(n)}$ comes from the imaginary part of the one-loop contribution
($C_F=4/3$)
\begin{eqnarray}
H_I^{1(n)}({\it loop\/})&=&2N_cs\frac{\alpha_sC_F}{4\pi}\pi v
  \frac{m_t}{\sqrt{2s}}=H_I^{2(n)}(loop)\,,\label{loopI}\\
H_A^{4(n)}({\it loop\/})&=&2N_cs\frac{\alpha_sC_F}{4\pi}\pi(2-v^2)
  \frac{m_t}{\sqrt{2s}}\,.
\label{loopA}
\end{eqnarray}
As already mentioned in the Introduction, the transverse and normal
polarization components can be seen to fall off with a power behaviour of
$(\sqrt s)^{-1}$ relative to the longitudinal polarization components.

The $\alpha_s$ corrections to the polarized structure functions
$H_a^{j(m)}=H_a^j(+s^m)-H_a^j(-s^m)$ and the unpolarized structure function
$H_a^j=H_a(+s^m)+H_a(-s^m)$ ($s^m$ is the polarization vector of the top
quark) are too lengthy to be listed here. They can be found in
Refs.~\cite{gkt97,kpt94,gkt96,gk96}, or, in a very compact two-page
analytical representation, in Sec.~5 of Ref.~\cite{Groote:2008ux}.
 
\begin{figure}
\begin{center}
\epsfig{figure=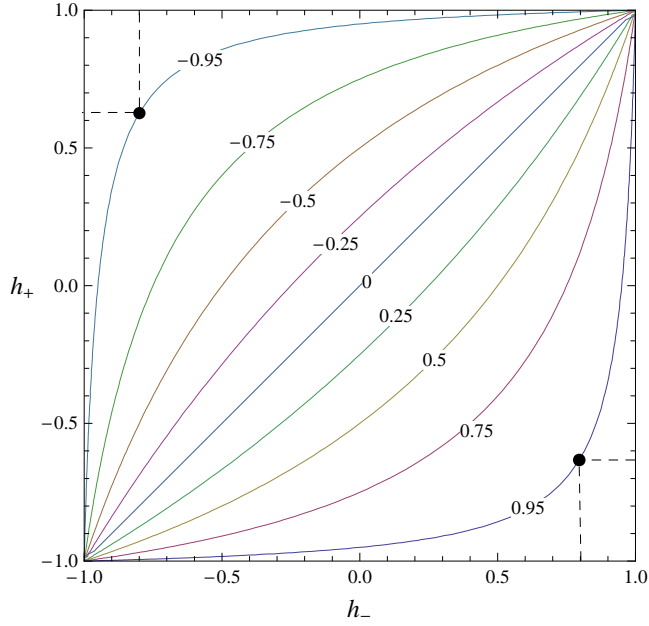, scale=1.0}\\[12pt]\qquad(a)\\[12pt]
\epsfig{figure=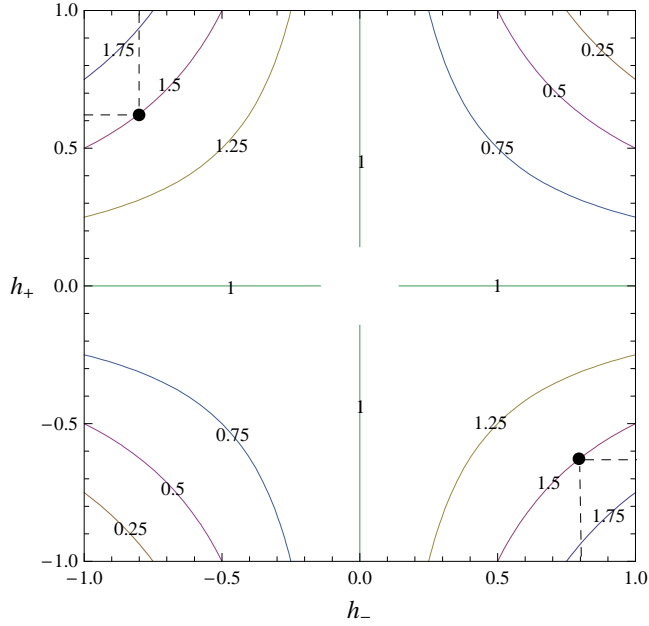, scale=1.0}\\[12pt]\qquad(b)
\end{center}
\caption{Contour plots in the $(h_-,h_+)$-plane (a) for fixed values of the
  effective polarization $P_{\rm eff}=(h_--h_+)/(1-h_-h_+)$ and (b) for fixed
  values of the gain factor $K_G=(1-h_-h_+)$. The two points marked off in the
  plots correspond to $(h_-,h_+)=(-0.8,+0.625)$ and $(+0.8,-0.625)$,
  respectively.
\label{f:fig3}}
\end{figure}

The longitudinal polarization of the electron and positron beams enter the
above formulas as~\cite{gkt97}\footnote{Transverse beam polarization effects
will not be discussed in this paper because present plans call for longitudinal
beam polarization at the ILC. Transverse beam polarization effects can be
included as described e.g.\ in Ref.~\cite{gkt97}.}
\begin{eqnarray}
\label{polrepl}
g_{1j}&\rightarrow&(1-h_-h_+)g_{1j}+(h_--h_+)g_{4j}
  \ =\ (1-h_-h_+)\big(g_{1j}+P_{\rm eff}g_{4j}\big)\,,\nonumber\\
g_{4j}&\rightarrow&(1-h_-h_+)g_{4j}+(h_--h_+)g_{1j}
  \ =\ (1-h_-h_+)\big(g_{4j}+P_{\rm eff}g_{1j}\big)\,,
\end{eqnarray}
where $P_{\rm eff}$ is defined in Eq.~(\ref{peff}). In Eq.~(\ref{polrepl})
$h_-$ denotes the electron's and $h_+$ denotes the positron's longitudinal
polarization which can take values between $\pm 1$. An electron with $h_-=\mp1$
will be referred to as the totally polarized left-handed (right-handed)
electron ($e^-_{L/R}$). Similarly, a right-handed positron ($e^+_R$) has
$h_+=-1$ and a left-handed positron ($e^+_L$) has $h_+=+1$. From the
definition of $P_{\rm eff}$ (see Eq.~(\ref{peff})) it is clear that large
values of $P_{\rm eff}$ can be reached even for nonmaximal values of $h_-$ and
$h_+$, as Fig.~\ref{f:fig3}a shows. For example, the large value of
$P_{\rm eff}=-0.95$ can be achieved with $h_-=-0.8$; $h_+=0.625$, and
correspondingly, $P_{\rm eff}=0.95$ can be reached with $h_-=0.8$;
$h_+=-0.625$. These two examples have been marked off in Fig.~\ref{f:fig3}.
Both sets correspond to a gain factor of $K_G=1.5$.

The orientation-dependent longitudinal, transverse, and normal polarization
components which we are interested in are defined by 
\begin{eqnarray}
\label{polnorm}
P^{(m)}(\cos\theta)=\frac{d\sigma^{(m)}/d\cos\theta}{d\sigma/d\cos\theta}
\qquad m=\ell,tr,n\,,
\end{eqnarray}
where $d\sigma/d\cos\theta$ is the unpolarized differential cross section.
Of course, there is an additional dependence of the above quantities on the
c.m.\ beam energy $\sqrt s$, and on the beam polarizations $h_-$ and $h_+$ to
be discussed later on. The unpolarized cross section is given by the first
three terms in Eq.~(\ref{diffcross}) dropping, of course, the label $(\ell)$.

Dropping the common factor $\pi\alpha^{2}v/(3s^{2})$ in the ratio
(\ref{polnorm}), we shall represent the polarization components by the ratios
\begin{equation}
\label{NoverD}
P^{(m)}(\cos\theta)=\frac{N^{(m)}(\cos\theta)}{D(\cos\theta)}\,, 
\qquad m=\ell,tr,n.
\end{equation}
In particular, the gain factor $K_G$ has canceled out in the ratio
(\ref{NoverD}) implying that the polarization only depends on $P_{\rm eff}$.

The numerator factors $N^{(m)}$ in Eq.~(\ref{NoverD}) are given by
\begin{eqnarray}
\label{lpol}
N^{(\ell)}(\cos\theta)&=&\frac38(1+\cos^2\theta)\,
  (g_{14}+g_{44}P_{\rm eff})H_U^{4(\ell)}
  +\frac34\sin^2\theta\,(g_{14}+g_{44}P_{\rm eff})H_L^{4(\ell)}\nonumber\\&&
  +\frac34\cos\theta\,\bigg((g_{41}+g_{11}P_{\rm eff})H_F^{1(\ell)}
  +(g_{42}+g_{12}P_{\rm eff})H_F^{2(\ell)}\bigg)\,,
\end{eqnarray}
\begin{eqnarray}
\label{trpol}
N^{(tr)}(\cos\theta)\!\!\!&=&\!\!\!-\frac3{\sqrt2}\sin\theta\cos\theta\,\,
(g_{14}+g_{44}P_{\rm eff})\,H_I^{4(tr)}\nonumber\\&&
  -\frac3{\sqrt2}\sin\theta\left((g_{41}+g_{11}P_{\rm eff})H_A^{1(tr)}
  +(g_{42}+g_{12}P_{\rm eff})H_A^{2(tr)}\right)\,,
\end{eqnarray}
and by
\begin{eqnarray}
N^{(n)}(\cos\theta)\!\!\!&=&\!\!\!-\frac3{\sqrt2}\sin\theta\cos\theta\,
  \left((g_{11}+g_{41}P_{\rm eff})\,H_{I}^{1(n)}({\it loop\/})
  +(g_{12}+g_{42}P_{\rm eff})\,H_{I}^{2(n)}({\it loop\/})\right)\nonumber\\&&
  -\frac3{\sqrt2}\sin\theta\,(g_{44}+g_{14}P_{\rm eff})\,
  H_A^{4(n)}({\it loop\/})\,.
\end{eqnarray}
For the denominator, one has
\begin{eqnarray}
\label{D}
D(\cos\theta) &=&\frac38(1+\cos^2\theta)\,\bigg((g_{11}+g_{41}P_{\rm eff})
  H_U^1+(g_{12}+g_{42}P_{\rm eff})H_U^2\bigg)\nonumber\\&&
  +\frac34\sin^2\theta\,\bigg((g_{11}+g_{41}P_{\rm eff})H_L^1
  +(g_{12}+g_{42}P_{\rm eff})H_L^2\bigg)\nonumber\\&&
  +\frac34\cos\theta\,(g_{44}+g_{14}P_{\rm eff})H_{F}^{4}\,\,.
\end{eqnarray}

At the forward (FP) and backward (BP) point the transverse and normal
polarization components vanish. Referring to the relations~(\ref{identities}),
at Born term level the longitudinal polarization component $P^{(\ell)}$ takes
a very simple form at the forward (FP) and backward (BP) point for the maximal
values of the effective polarization $P_{\rm eff}=\pm 1$. One has
\begin{eqnarray}
\label{fpbp}
FP:\qquad P^{(\ell)}(\cos\theta=+1)&=&\pm 1\,,\nonumber\\
BP:\qquad P^{(\ell)}(\cos\theta=-1)&=&\mp 1\,,
\end{eqnarray}
in agreement with angular momentum conservation. It is clear that these 
relations no longer hold true in general at NLO due to hard gluon emission. 

It is useful to define the left--right polarization asymmetry $A_{LR}$ through
the relation
\begin{equation}
\frac{d\sigma(P_{\rm eff})-d\sigma(-P_{\rm eff})}
{d\sigma(P_{\rm eff})+d\sigma(-P_{\rm eff})}=- A_{LR}P_{\rm eff}\,,
\end{equation}
where
\begin{equation}
\label{LRasym}
A_{LR}=-\frac{\frac38(1+\cos^2\theta)\,\Big(g_{41}H_U^1+g_{42}H_U^2\Big)
  +\frac34\sin^2\theta\,\Big(g_{41}H_L^1+g_{42}H_L^2\Big)
  +\frac34\cos\theta g_{14}H_F^4}
  {\frac38(1+\cos^2\theta)\,\bigg(g_{11}H_U^1+g_{12}H_U^2\bigg)
  +\frac34\sin^2\theta\,\bigg(g_{11}H_L^1+g_{12}H_L^2\bigg)
  +\frac34\cos\theta g_{44}H_F^4}
\end{equation}

Of interest is the angle $\alpha$ enclosed by the momentum and the
polarization of the top quark projected onto the scattering plane (see
Fig.~\ref{f:fig2}).\footnote{For the present purposes we neglect the
$O(\alpha_{s})$ normal component of the polarization vector which is quite
small. Note that, in general, one needs two angles to describe the orientation
of the polarization vector instead of the one angle $\alpha$ defined in
Eq.~(\ref{tan0}).} 
The angle $\alpha$ is determined by
\begin{equation}
\label{tan0}
\tan\alpha\,(\cos\theta)=\frac{N^{(tr)}(\cos\theta)}{N^{(\ell)}(\cos\theta)}\,.
\end{equation}
Equation~(\ref{tan0}) assumes a simple form at threshold and in the
high-energy limit as discussed in Sec.~4, and for $P_{\rm eff}=\pm1$ as will
be  discussed in Sec.~5. The correlations between $\alpha$ and $\theta$
implied by Eq.~(\ref{tan0}) will be discussed in Secs.~4, 5 and~7.

\section{Beam polarization dependence of the rate}

We begin our numerical discussion with the rate proportional to the denominator
expression in Eq.~(\ref{NoverD}). The effect of longitudinally polarized beams
on the polar averaged rate (called total rate) can be obtained from the form
\begin{equation}
\label{polsignum1}
\sigma=\sigma(P_{\rm eff}=0)\,\,(1-h_-h_+)\left(1+P_{\rm eff}
  \frac{g_{41}}{g_{11}}\frac{\displaystyle 1+\frac{g_{42}}{g_{41}}
  \frac{H_{U+L}^2}{H_{U+L}^1}}{\displaystyle 1+\frac{g_{12}}{g_{11}}
  \frac{H_{U+L}^2}{H_{U+L}^1}}\right)\,,
\end{equation}
which, at the Born level and at $\sqrt s=500\GeV$, gives
\begin{equation}
\label{polsignum2}
\sigma=\sigma(P_{\rm eff}=0)\,\,(1-h_-h_+)\Big(1-0.37\,P_{\rm eff}\Big)\,.
\end{equation}

From Eq.~(\ref{polsignum2}) it is evident that the total rate becomes maximal
on two counts: (i) large values of the gain factor $K_G=(1-h_-h_+)$, requiring
$sign(h_-)=-sign(h_+)$; and (ii) large negative values of $P_{\rm eff}$,
which can be achieved with large negative and positive values of $h_-$ and
$h_+$, respectively. The maximal enhancement of the rate will be obtained for
$h_-=-1$ and $h_+=+1$ such that $P_{\rm eff}=-1$ and $K_G=2$. At
$\sqrt s=500\GeV$, this leads to a maximal enhancement factor of $2.74$ over
the unpolarized case. It is interesting to note that for $(b\bar b)$
production at $\sqrt s=500\GeV$ the effective enhancement through beam
polarization effects is slightly larger than in the $(t\bar t)$ case. For
$(b\bar b)$ production the last factor in Eq.~(\ref{polsignum1}) is replaced
by the simpler expression $(1+P_{\rm eff}\,g_{41}/g_{11})$ since, at
$\sqrt s=500\GeV$, the ratio $H_{U+L}^2/H_{U+L}^1$ is practically zero for
bottom quark production. Using the results of the Appendix applied to the
$(b\bar b)$ case, one finds $g_{41}/g_{11}=-0.62$ leading to an overall
enhancement factor of $3.24$ for the optimal choice of parameters $h_-=-1$ and
$h_+=+1$ ($P_{\rm eff}=-1$) at $\sqrt s=500\GeV$.

In Fig.~\ref{f:fig3}b we show some contour lines for fixed values of the gain
factor $K_G=(1-h_-h_+)$ in the $(h_-,h_+)$-plane. Clearly, quadrants 2 and 4
are favoured if one wants to obtain a gain factor exceeding one, i.e.\
$K_G\ge 1$. As concerns the rate dependence on $P_{\rm eff}$ (rightmost factor
in Eq.~(\ref{polsignum2})), a further rate enhancement is achieved for negative
values of $P_{\rm eff}$, i.e. one would have to choose points lying to the
left of the line $h_-=-h_+$ in Fig.~\ref{f:fig3}a. The optimal choice as
concerns the rate would thus be quadrant 2 in the $(h_-,h_+)$-plane. One notes
that large negative values of $P_{\rm eff}$ can readily be achieved for
nonmaximal values of the beam polarization, as illustrated in
Fig.~\ref{f:fig3}a, where we have plotted some contour lines in the
$(h_-,h_+)$ plane corresponding to fixed values of $P_{\rm eff}$. One notes
that the regions of large $K_G$ and large negative $P_{\rm eff}$ have a large
overlap. We mention that one may have to give up the optimal choice in the
$(h_-,h_+)$ plane if one wants to achieve other goals such as minimizing the
polarization. 
\begin{figure}
\begin{center}
\epsfig{figure=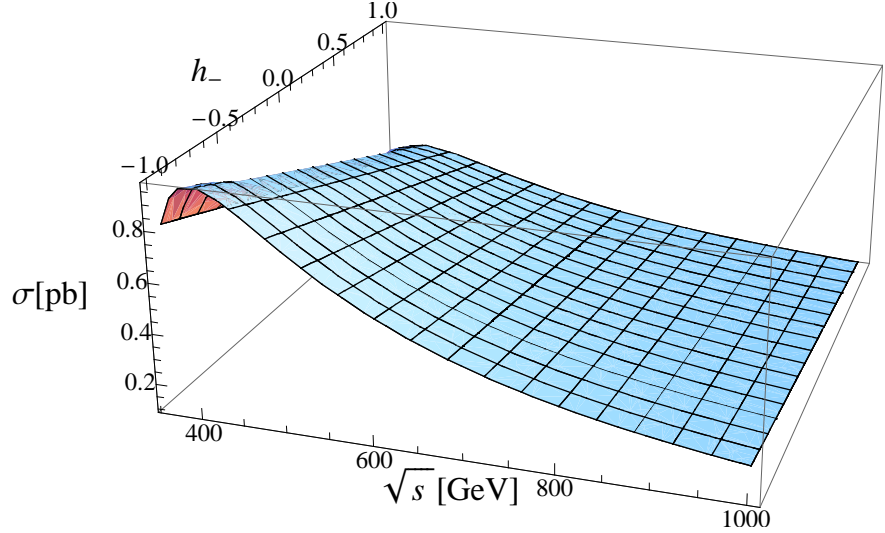, scale=1.3}\\[12pt]\qquad(a)\\[12pt]
\epsfig{figure=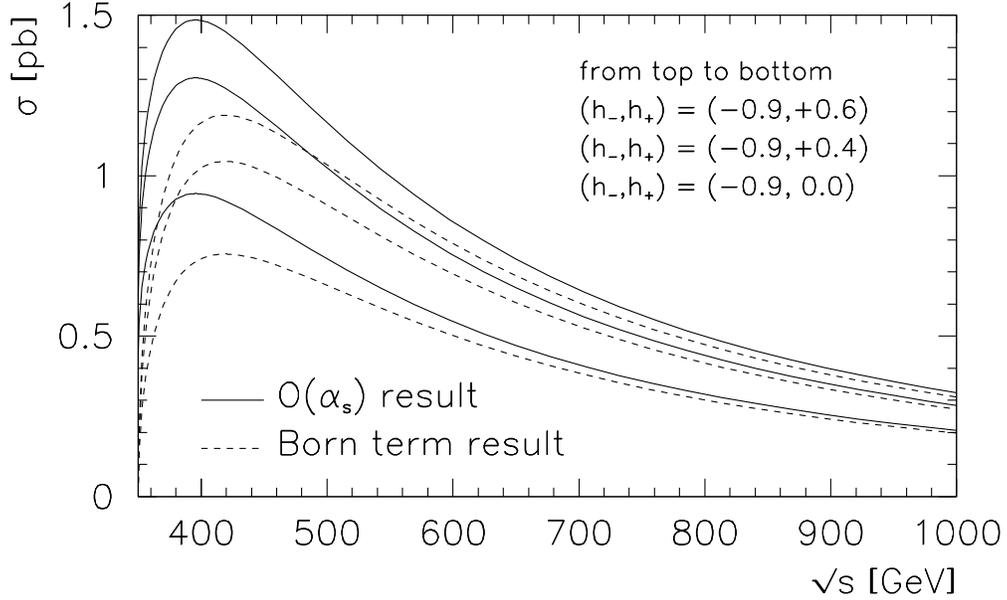, scale=0.8}\\[12pt]\qquad(b)
\end{center}
\caption{The total cross section $\sigma$ at the one-loop level as a function
of the beam energy $\sqrt s$ and (a) the electron polarization $h_-$
($h_+=0)$; (b) for three values of the positron polarization $h_+=0$,
$0.4$, $0.6$, and with the fixed electron beam polarization of $h_-=-0.9$
(solid lines). In Fig.~\ref{f:fig4}b we also show the respective LO rates
(dashed lines).
\label{f:fig4}}
\end{figure}

The QCD one-loop corrections to the total cross section are well-known (see
e.g.\ Ref.~\cite{krz86}) and add about $13\%$ at $\sqrt s=500\GeV$ to the Born
total cross section, where the percentage increase has very little dependence
on the beam polarization. We mention that the electroweak corrections to the
total rate are smaller, and amount to about $50\%$ of the QCD
corrections~\cite{Kuhn:1999kp}. For the strong coupling $\alpha_s$ we
use two-loop running adjusted to the value $\alpha_s(m_Z)=0.1175$ and fitted at
$2m_t=350\GeV$.\footnote{For $\alpha$ we take the value $\alpha=1/137$. If one
uses a running $\alpha$, for example $\alpha=1/128$, the cross sections in
Fig.~\ref{f:fig4} would increase by $14.6\%$.} Close to threshold the
$O(\alpha_s)$ corrections become larger and amount to about $27\%$ of the
total cross section at e.g.\ $\sqrt s=400\GeV$. The c.m.\ energy dependence of
the total cross section $\sigma$ is shown in Figs.~\ref{f:fig4}a
and~\ref{f:fig4}b. In Fig.~\ref{f:fig4}a we take $h_+=0$ and show the energy
dependence of the total cross section varying $h_-$ over its whole range
$[-1,+1]$. One notes a strong dependence on $h_-$ apart from the standard
falloff of the total cross section with beam energy. Since for $h_+=0$ the
gain factor $K_G$ is equal to $1$ and since $P_{\rm eff}=h_-$, the rate
depends linearly on $h_-$ as displayed in Eqs.~(\ref{polsignum1})
and~(\ref{polsignum2}). The rate is largest for $h_-=-1$ and then linearly
drops to its lowest value at $h_-=+1$. In Fig.~\ref{f:fig4}b we show the
energy dependence of the rate for the three pairs of beam polarizations
$(h_-,h_+)=(-0.9,0),(-0.9,+0.4),(-0.9,+0.6)$. If one translates this into the
$(K_G,P_{\rm eff})$ representation, one has $(K_G,P_{\rm eff})=(1,-0.9),
(1.36,-0.96),(1.54,-0.97)$. The hierarchy of rates in Fig.~\ref{f:fig4}b can
be seen to be mostly determined by the gain factor $K_G$ in
Eq.~(\ref{polsignum1}).

\begin{figure}
\begin{center}
\epsfig{figure=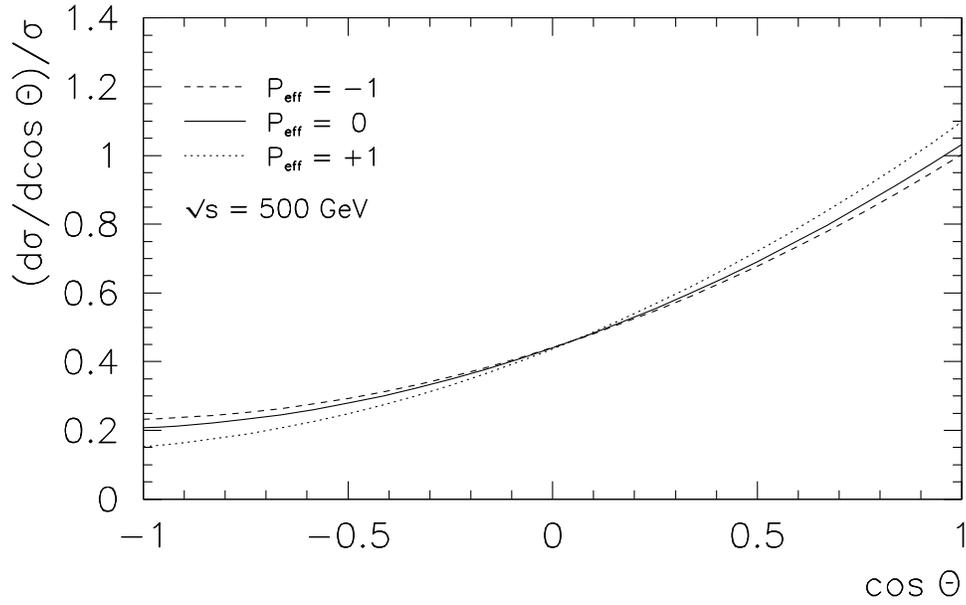, scale=0.8}\\[12pt]\qquad(a)\\[12pt]
\epsfig{figure=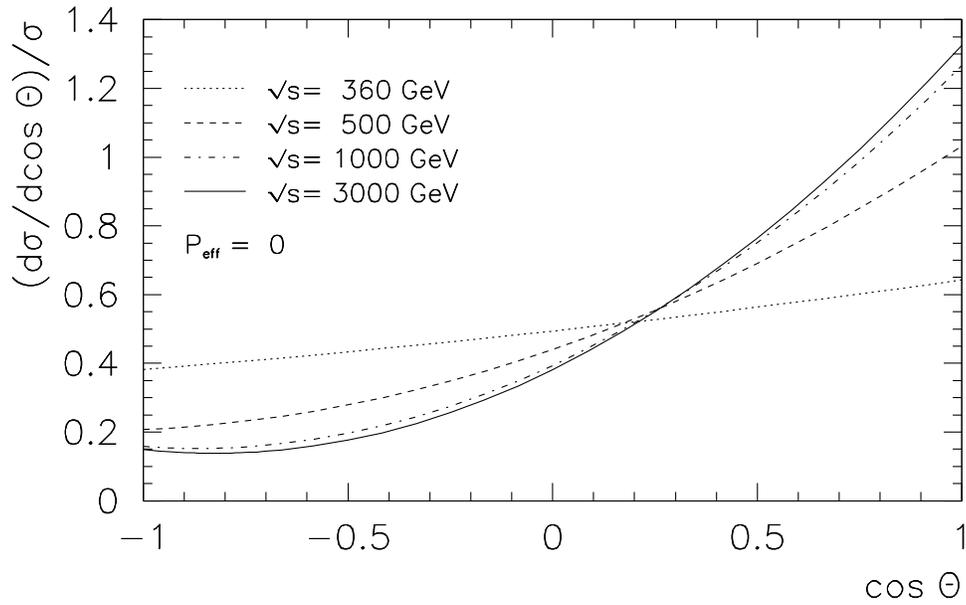, scale=0.8}\\[12pt]\qquad(b)
\end{center}
\caption{Polar angle dependence of the differential cross section for (a)
  $\sqrt s=500\GeV$ and $P_{\rm eff}=-1,0,+1$ and (b) $P_{\rm eff}=0$ for
  beam energies $\sqrt{s}=360\GeV$ (dotted line), $500\GeV$ (dashed line),
  $1000\GeV$ (dash-dotted line), and $3000\GeV$ (solid line)
\label{sigmar}}
\end{figure}  

\begin{figure}[t]
\begin{center}
\epsfig{figure=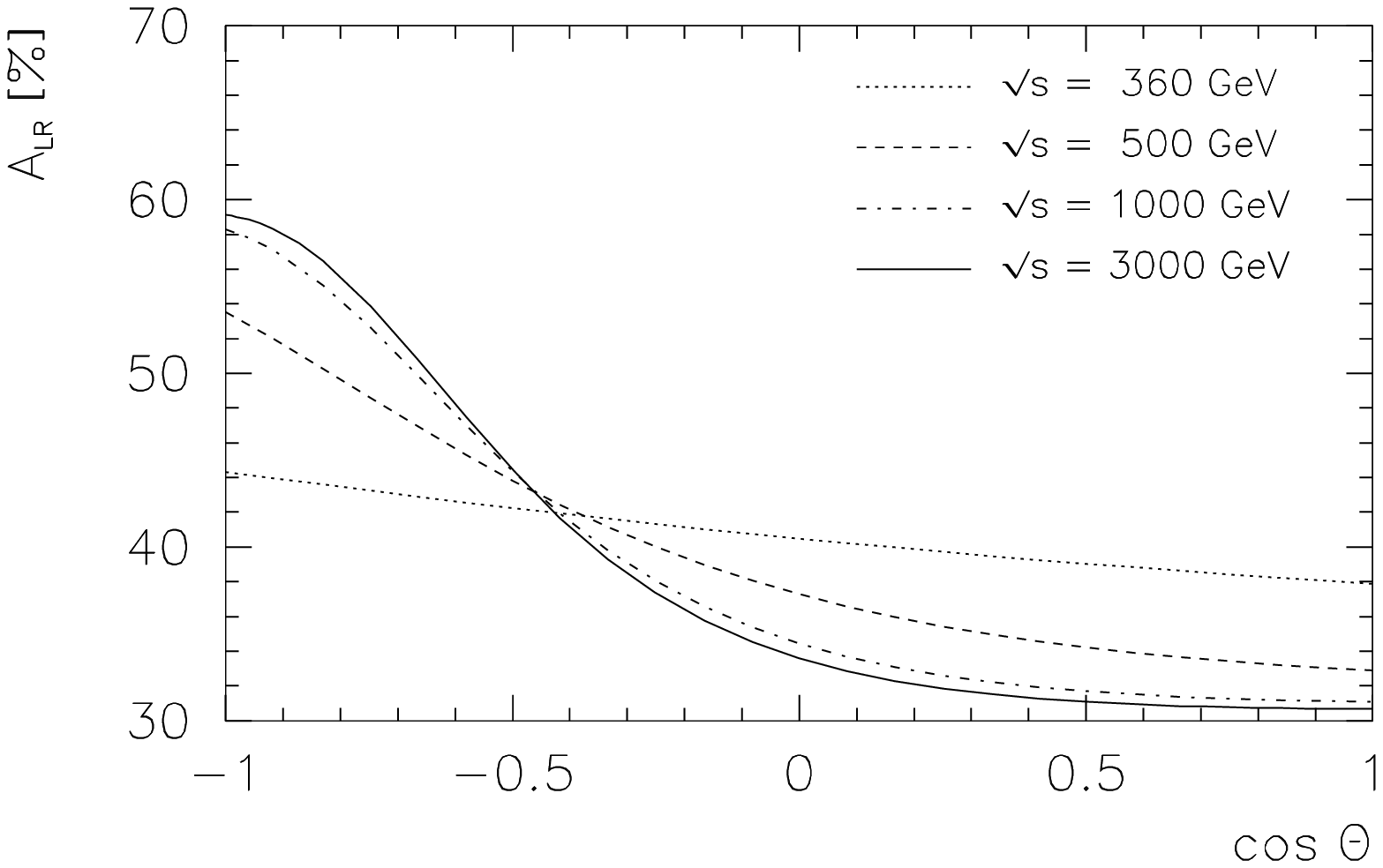, scale=0.8}
\end{center}
\caption{Left--right polarization asymmetry $A_{LR}$ for  
  $\sqrt s=360$, $500$, $1000$, and $3000\GeV$ (notation as in
  Fig.~\ref{sigmar})
  \label{f:asymrl}}
\end{figure}

Next we turn to the differential rate distribution with respect to
$\cos\theta$. In order to illustrate the forward dominance of the differential
$\cos\theta$-distribution we plot $\sigma^{-1}d\sigma/d\cos\theta$ against
$\cos\theta$. Note that the dependence on the gain factor $K_G=1-h_-h_+$ drops
out in the ratio. In Fig.~\ref{sigmar}a we plot the differential rate
distribution for a fixed value of $\sqrt s=500\GeV$ and for
$P_{\rm eff}=-1,0,+1$. One sees a pronounced forward dominance of the
differential distribution which does not depend much on the value of
$P_{\rm eff}$. In Fig.~\ref{sigmar}b we keep the effective beam polarization
fixed at $P_{\rm eff}=0$ and vary $\sqrt s$ through several values. At
threshold $\sqrt s=350\GeV$ one has a flat distribution
$\sigma^{-1}d\sigma/d\cos\theta=0.5$. When the energy is increased, the forward
rate clearly dominates over the backward rate. The forward dominance becomes
even stronger for increasing energies.

Of related interest is the rate into the forward (F) and backward (B)
hemispheres. Again, the gain factor $K_G$ drops out in the ratio. At
$\sqrt s=500\GeV$, one numerically obtains 
\begin{equation}
\label{sigmaFB}
\frac{\langle\sigma\rangle_F}{\langle\sigma\rangle_B}
  =\frac{\langle\sigma\rangle_F}{\langle\sigma\rangle_B}\bigg|_{P_{\rm eff}=0}
  \,\,\,\,
\frac{1-0.34\,P_{\rm eff}}{1-0.43\,P_{\rm eff}}
  =\left\{\begin{array}{rl}
  +2.73&\quad P_{\rm eff}=+1\\
  +2.36&\quad\phantom{P_{\rm eff}}=\phantom{+}0\\
  +2.21&\quad\phantom{P_{\rm eff}}=-1
\end{array}\right\}\,.
\end{equation}
The mean forward rate $\langle\sigma\rangle_F$ clearly dominates over the mean
backward rate $\langle\sigma\rangle_B$. The dependence of the $F/B$ rate ratio
on $P_{\rm eff}$ is not very pronounced.

In Fig.~\ref{f:asymrl} we plot the polar angle dependence of the NLO
left-right polarization asymmetry $A_{LR}$ for different energies. At
$\sqrt s=360$ GeV the $\cos\theta$ dependence already starts to deviate from
the flat Born term behaviour at threshold given by
$A_{LR}=-(g_{41}+g_{42})/(g_{11}+g_{12})=0.409$. The left-right polarization 
asymmetry $A_{LR}$ peaks toward the backward region and reaches $\approx 59\%$
at the backward point for the highest energy $\sqrt s=3000\GeV$ in
Fig.~\ref{f:asymrl}.

\section{Born term simplifications at threshold\\ and in the high-energy limit}

Before turning to the numerical analysis of the polarization of the top quark,
in this section we shall first discuss Born term simplifications of the
polarization of the top quark at threshold and in the high-energy limit. In
Sec.~5 we discuss Born term simplifications that occur for $P_{eff}=\mp1$.

At threshold $v\to 0$ and in the high-energy limit $v\to 1$, the polarization
expressions become quite simple. At threshold, the polarization of the top
quark is parallel to the beam axis, regardless of the polar orientation of the
top quark (see e.g.\ Ref.~\cite{Tsai:1971vv}). In fact, a large part of the
beam polarization gets transferred to the polarization of the top quark at
threshold. For the Born term contributions the top quark polarization at
threshold can be calculated from Eqs.~(\ref{lpol}), (\ref{trpol})
and~(\ref{D}) (see also Ref.~\cite{krz86,k01}). It is nominally given
by\footnote{As discussed in Sec.~2, QCD binding effects significantly modify
the naive threshold results in the threshold region.} 
\begin{equation}
\label{thresh}
\vec{P}=\frac{P_{\rm eff}- A_{LR}}{1-P_{\rm eff}A_{LR}}\,\,\,\hat{n}_{e^-}\,,
\end{equation}
where $A_{LR}$ is the left-right beam polarization asymmetry
$(\sigma_{LR}-\sigma_{RL})/(\sigma_{LR}+\sigma_{RL})$ at threshold
(see Eq.~(\ref{LRasym})) and $\hat{n}_{e^-}$ is a unit vector pointing into
the direction of the electron momentum. In terms of the electroweak coupling
parameters $g_{ij}$ (see the Appendix), the nominal polarization asymmetry at
threshold $\sqrt s=2m_t$ is given by
$A_{LR}=-(g_{41}+g_{42})/(g_{11}+g_{12})=0.409$. The simplification at
threshold arises from the fact that, from the four $(L,S)_{V,A}$ amplitudes
$(L,S)_{V,A}=(0,1)_V,(2,1)_V,(1,0)_A,(1,1)_A$ describing the production of a
spin-1/2 pair, only the $S$-wave amplitude $(0,1)_V$ survives at threshold.
The suffices $V$ and $A$ denote vector current (V) and axial vector current (A)
production. Correspondingly, the combinations $(g_{41}+g_{42})$ and
$(g_{11}+g_{12})$ contain only the vector current coupling on the quark side.

The magnitude of the threshold polarization is given by
\begin{equation}
\label{thrpol}
|\vec{P}\,|=\left|\frac{P_{\rm eff}-A_{LR}}{1-P_{\rm eff}A_{LR}}\right|\,.
\end{equation}
The threshold polarization is independent of $\cos\theta$, i.e.\
$\langle|\vec{P}\,|\rangle=|\vec{P}\,|$. The polarization vanishes for
$P_{\rm eff}= A_{LR}$ independent of $\cos\theta$.\footnote{Threshold
simplifications for $(q \bar q)$ production have also been discussed in
Ref.~\cite{Groote:2009dd}. Similar simplifications for polarization
observables occur for the threshold production of gauge boson
pairs~\cite{Groote:2010nk}.} For $P_{\rm eff}>A_{LR}$
and $P_{\rm eff}< A_{LR}$ one has $\vec{P}=|\vec{P}\,|\,\hat{n}_{e^-}$ and
$\vec{P}=-|\vec{P}\,|\,\hat{n}_{e^-}$, respectively, such that
$P^{(tr)}=\mp|\vec{P}\,|\sin\theta$ and
$P^{(\ell)}=\pm|\vec{P}\,|\cos\theta$. In particular, one has a $100\%$
threshold polarization of the top quark for $P_{\rm eff}=\pm1$ with
$\vec{P}=\pm\hat{n}_{e^-}$.

Extrapolations away from $P_{\rm eff}=\pm 1$ are more stable for
$P_{\rm eff}=-1$ than for $P_{\rm eff}=+1$ as the slope of Eq.~(\ref{thrpol})
at $P_{\rm eff}=\pm 1$ shows. One has
\begin{equation}
\label{slope}
\frac{d|\vec{P}\,|}{dP_{\rm eff}}=\pm\frac{1\pm A_{LR}}{1\mp A_{LR}}.
\end{equation}
For $P_{\rm eff}=-1$ one has a slope of $-(1-A_{LR})/(1+A_{LR})=-0.42$ while
one has a much larger positive slope of $(1+A_{LR})/(1-A_{LR})=+2.38$ for
$P_{\rm eff}=+1$. This substantiates the statement made above and in Sec.~1
about the stability of extrapolations away from $P_{\rm eff}=\pm 1$. For
example, keeping only the linear term in the Taylor expansion of
Eq.~(\ref{thrpol}), one has $|\vec{P}\,|=0.98$ for $P_{\rm eff}=-0.95$, while
$|\vec{P}\,|$ drops to $|\vec{P}\,|=0.88$ for $P_{\rm eff}=+0.95$.

For energies above threshold the slope Eq.~(\ref{slope}) becomes energy and
angle dependent. We do not show plots of the slope at higher energies. We have,
however, checked numerically that the above statement about the stability of
the $|\vec{P}\,|$ result at $P_{\rm eff}=-1$ against variations of
$P_{\rm eff}$ remains true at higher energies in the whole angular range,
where the slope in the backward region has a tendency to be smaller than in
the forward region.

As mentioned above, minimal polarization $|\vec{P}\,|=0$ occurs for
$P_{\rm eff}= A_{LR}=0.409$ for all values of $\cos\theta$. This again shows
that an extrapolation away from $P_{\rm eff}=-1$ is more stable than an
extrapolation from $P_{\rm eff}=+1$ since one is much closer to the
polarization zero in the latter case. This observation will carry over to the
$P_{\rm eff}$-dependence at higher energies.

\begin{figure}
\begin{center}
\epsfig{figure=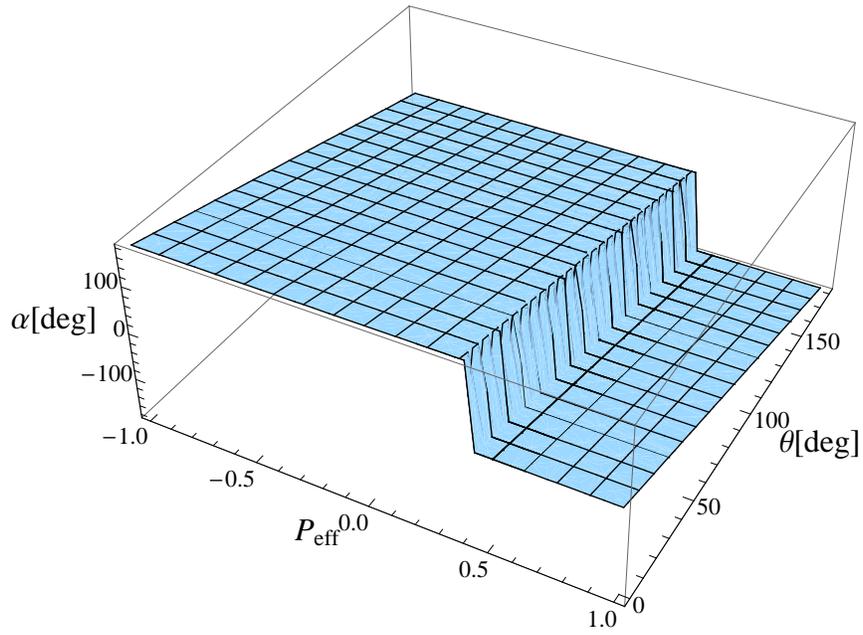, scale=1.0}\\[12pt]\qquad(a)\\[12pt]
\epsfig{figure=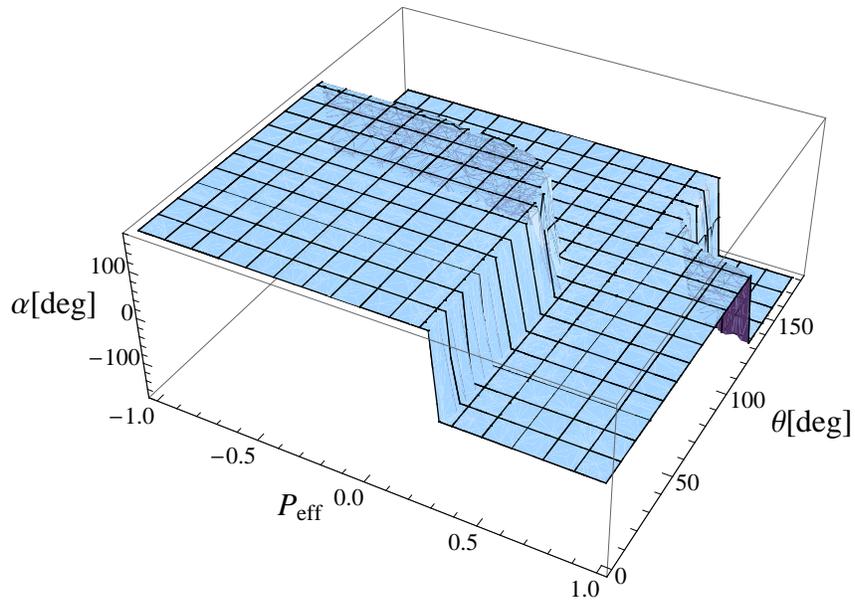, scale=1.0}\\[12pt]\qquad(b)
\end{center}
\caption{Correlation of the angles $\alpha$ and $\theta$ in dependence on the
  effective beam polarization $P_{\rm eff}$ (a) for threshold energies
  $\sqrt s=2m_t$ and (b) for $s\to\infty$
\label{f:fig7}}
\end{figure}

In Fig.~\ref{f:fig7}a we show the threshold correlation of the angles $\alpha$
and $\theta$ for different values of $P_{\rm eff}$. Starting at
$P_{\rm eff}=-1$ the two angles are related by $\alpha=180^{\circ}-\theta$ up
to the longitudinal polarization zero at $P_{\rm eff}=A_{LR}=0.409$ after which
the correlation becomes $\alpha=-\theta$.

As the beam energy increases, the polarization vector of the top quark slowly
turns into the direction of its momentum (or opposite to it). Finally, in the
high-energy limit $s\rightarrow\infty$, when $v\to 1$, the polarization of the
top becomes purely longitudinal in the helicity system such that
$|\vec{P}\,|=|P^{(\ell)}|$ since its transverse and normal components
involve a spin flip amplitude and thus vanish as $m_t/\sqrt s$. Note that,
although $P^{(tr)}$ is asymptotically suppressed, it is still sizable at
$\sqrt s=1000\GeV$ as Fig.~\ref{f:fig1} shows.

In fact, in the high-energy limit, one has
$\vec{P}(\cos\theta)=P^{(\ell)}(\cos\theta)\cdot\hat{p_t}$ with
\begin{eqnarray}
\label{helimit}
P^{(\ell)}(\cos\theta)&=&\nonumber\\&&\hspace*{-2.3cm}
\frac{(g_{14}+g_{41}+P_{\rm eff}(g_{11}+g_{44}))(1+\cos\theta)^2
  +(g_{14}-g_{41}-P_{\rm eff}(g_{11}-g_{44}))(1-\cos\theta)^2}
  {(g_{11}+g_{44}+P_{\rm eff}(g_{14}+g_{41}))(1+\cos\theta)^2
  +(g_{11}-g_{44}-P_{\rm eff}(g_{14}-g_{41}))(1-\cos\theta)^2}\nonumber\\
\end{eqnarray}
for the surviving longitudinal polarization. In the same limit, the electroweak
coupling coefficients take the numerical values $g_{11}=0.601$,
$g_{14}=-0.131$, $g_{41}=-0.201$, $g_{44}=0.483$, $g_{12}=0.352$, and
$g_{42}=-0.164$. When $P_{\rm eff}=-1$ it is more convenient to switch to the
chiral electroweak coefficients $f_{LL/LR}$ defined in the Appendix. One has
($f_{LL}=-1.190$; $f_{LR}=-0.434$)
\begin{equation}
\label{Lhelimit}
P^{(\ell)}(\cos\theta)=-\frac{1-b_{LR}}{1+b_{LR}}\qquad\mbox{with}\qquad
b_{LR}=\left(\frac{f_{LR}}{f_{LL}}\right)^2
\frac{(1-\cos\theta)^2}{(1+\cos\theta)^2}\,.
\end{equation}
$P^{(\ell)}$ goes through zero for $b_{LR}=1$ which is solved by
$\cos\theta=-(f_{LL}-f_{LR})/(f_{LL}+f_{LR})$. For $P_{\rm eff}=+1$ one has a
similar simplification where the quantities on the right-hand side of
Eq.~(\ref{Lhelimit}) are replaced by $b_{LR}\to b_{RL}$ and
$f_{LL/LR}\to f_{RR/RL}$ ($f_{RR}=-0.867$; $f_{RL}=-0.217$). In this case
$P^{(\ell)}$ goes through zero for $b_{RL}=1$, or for
$\cos\theta=-(f_{RR}-f_{RL})/(f_{RR}+f_{RL})$.

At threshold the rate shows no $\cos\theta$ dependence since the
$(t\bar t)$ pair is produced in a $S$-wave state. This is different in the
high-energy limit when $v=1$, where the forward rate strongly dominates over
the backward rate, as an inspection of the denominator of Eq.~(\ref{helimit})
shows. Since an accurate measurement of the polarization observables requires
large statistics, and thus large event samples, the issue of rates is an
important one. Numerically, one finds
$\sigma(\cos\theta=+1)/\sigma(\cos\theta=-1)
=9.23\,(1-0.31P_{\rm eff})/(1-0.60P_{\rm eff})$. The dependence on
$P_{\rm eff}$ is small. When averaging over the forward ($F$) and backward
($B$) hemispheres, one finds $\langle\sigma\rangle_F/\langle\sigma\rangle_B
=4.04\,(1-0.31P_{\rm eff})/(1-0.43P_{\rm eff})$, i.e.\ in the case of
unpolarized beams when $P_{\rm eff}=0$ the rate in the forward hemisphere
dominates over the rate in the backward hemisphere by a factor of four with
only slight dependence on beam polarization. Comparing to Eq.~(\ref{sigmaFB})
the forward dominance is more pronounced in the high-energy limit than at
$\sqrt s=500\GeV$.

Equation~(\ref{helimit}) also very nicely shows how varying $P_{\rm eff}$
affects the longitudinal polarization $P^{(\ell)}$. For the unpolarized beam
case $P_{\rm eff}=0$ the longitudinal polarization $P^{(\ell)}$ is negative
($-31\%$) at the forward point (FP) $\cos\theta=+1$ and positive ($+60\%$) at
the backward point (BP) $\cos\theta=-1$. For maximally polarized beams
$P^{\rm eff}=\pm1$, Eq.~(\ref{helimit}) can be seen to satisfy the angular
momentum conservation conditions, Eq.~(\ref{fpbp}). For $P^{\rm eff}=\pm1$ the
longitudinal polarization monotonically increases/decreases from the backward
to the forward point. It can be seen to go through zero at
$\cos\theta=(g_{11}-g_{41}-g_{12}+g_{42})/(g_{14}-g_{44})=-0.47\
(\,\simeq\,117.8^\circ)$ for $P_{\rm eff}=-1$ and
$\cos\theta=-(g_{11}+g_{41}-g_{12}-g_{42})/(g_{14}+g_{44})=-0.60\
(\,\simeq\,126.9^\circ)$ for $P_{\rm eff}=+1$ (see discussion after
Eq.~(\ref{Lhelimit})). Close to $P_{\rm eff}=\pm1$, the longitudinal
polarization zeros are only mildly dependent on $P_{\rm eff}$. There is a
range of $P_{\rm eff}$ values for which the longitudinal polarization remains
positive over the whole $\cos\theta$ range. This is determined by the zeros of
the coefficients of the angular factors in the numerator of
Eq.~(\ref{helimit}). The condition for positivity of $P^{(\ell)}$ reads
\begin{equation}
-\frac{g_{14}+g_{41}}{g_{11}+g_{44}}<P_{\rm eff}
<\frac{g_{14}-g_{41}}{g_{11}-g_{44}}\,.
\end{equation}
Numerically this translates into $0.31<P_{\rm eff}<0.60$. The same bounding
values determine the vanishing of the polarization at the forward and backward
points. At the forward point, where the rate is highest, the polarization
$|\vec{P}\,|$ can be made to vanish by setting
$P_{\rm eff}=-(g_{14}+g_{41})/(g_{11}+g_{44})=0.31$. At the backward point,
one has zero longitudinal polarization for
$P_{\rm eff}=(g_{14}-g_{41})/(g_{11}-g_{44})=0.60$.

All of this is illustrated in Fig.~\ref{f:fig7}b showing the correlation
between $P_{\rm eff}$ and the angles $\alpha$ and $\theta$. The steplike
behaviour in Fig.~\ref{f:fig7}b is associated with the vanishing of the
polarization at which points the polarization vector changes its direction by
$180^{\circ}$. At $P_{\rm eff}=-1$ the polarization vector $\vec{P}$ is
antiparallel to $\vec{p}_t$ up to where $\vec{P}$ becomes zero at
$\theta\sim117.8^{\circ}$.  From then on $\vec{P}$ is parallel to $\vec{p}_t$.
Zero polarization and the location of the step-like behaviour is slightly
$P_{\rm eff}$-dependent and is shifted to lower values of $\alpha$. For
$0.31<P_{\rm eff}<0.60$ the polarization $\vec{P}$ is always parallel to
$\vec{p}_t$. Finally, for $P_{\rm eff}=+1$ the polarization $\vec{P}$ starts
off parallel to $\vec{p}_t$ and turns antiparallel to $\vec{p}_t$ after the
zero at $\cos\theta\simeq126.9^{\circ}$. Again the polarization zero and the
associated step-like behaviour is slightly shifted when one moves away from
$P_{\rm eff}=+1$.

\begin{figure}
\begin{center}
\epsfig{figure=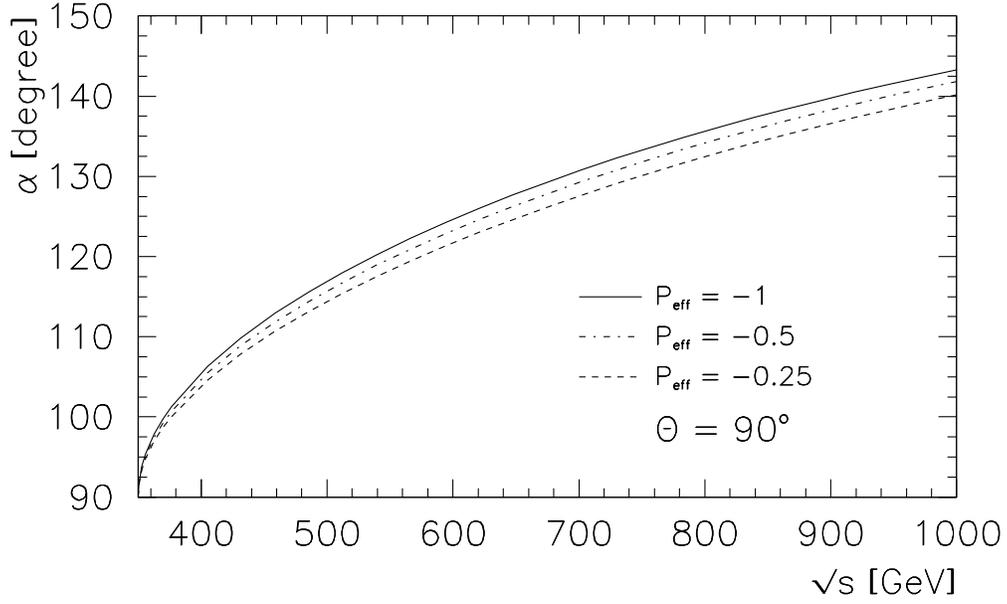, scale=0.8}\\[12pt]\qquad(a)\\[12pt]
\epsfig{figure=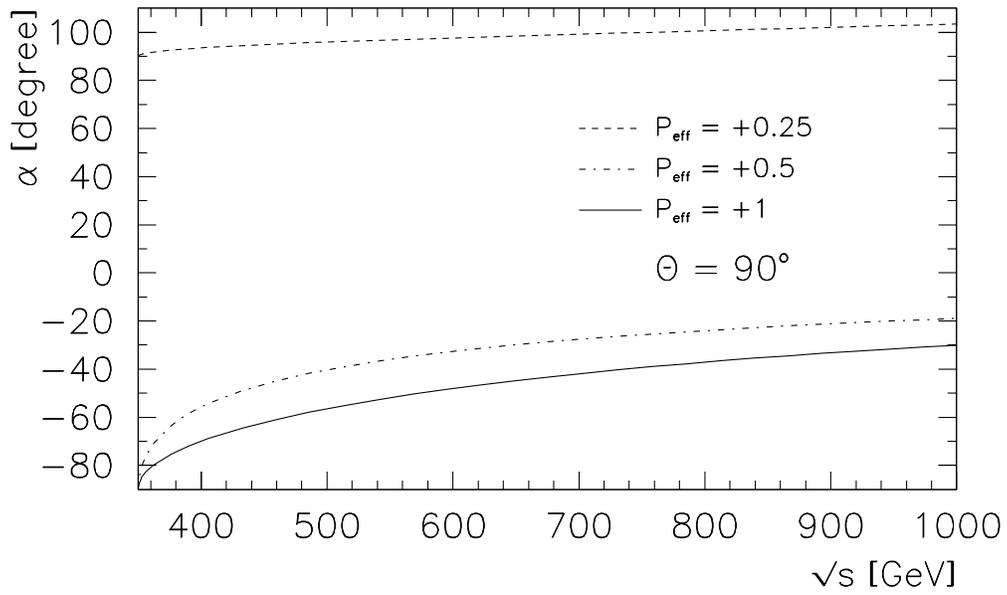, scale=0.8}\\[12pt]\qquad(b)
\end{center}
\caption{The top quark polarization angle $\alpha$ for a scattering angle of
  $\theta=90^\circ$ as a function of the beam energy for (a) negative
  values of $P_{\rm eff}=-1,-0.5,-0.25$ and (b) positive values
  of $P_{\rm eff}=+1,+0.5,+0.25$
  \label{f:fig8}}
\end{figure}

Given the fact that the polarization turns from the beam direction to the
momentum direction (or its opposite) going from threshold to the high energy
limit it would be interesting to know how fast this transition occurs when the
beam energy is ramped up in the envisaged range of beam energies
$\sqrt s\sim 2m_t\div 1000\GeV$. In Fig.~\ref{f:fig8} we investigate the
energy dependence of the angle $\alpha$ for several values of $P_{\rm eff}$
for a scattering angle of $\theta=90^\circ$. In Fig.~\ref{f:fig8}a we consider
three representative negative values of $P_{\rm eff}$. All three curves start
off with the threshold angle $\alpha=90^\circ$. The growth of $\alpha$ does
not depend much on $P_{\rm eff}$ but is still far away from the asymptotic
value $\alpha=180^\circ$ at $\sqrt s=1000\GeV$. For positive values of
$P_{\rm eff}$ the dependence of $\alpha$ on $P_{\rm eff}$ is more pronounced
(see Fig.~\ref{f:fig8}b). For $P_{\rm eff}=+1$ and $P_{\rm eff}=+0.5$, one is
getting closer to the asymptotic value of $\alpha=0^\circ$ at
$\sqrt s=1000\GeV$ than for the negative values of $P_{\rm eff}$ shown in
Fig.~\ref{f:fig8}a. The behaviour of the $P_{\rm eff}=+0.25$ curve differs from
the two other curves since one has crossed a longitudinal polarization zero
between $P_{\rm eff}=+0.5$ and $P_{\rm eff}=+0.25$. 

\section{Born term simplifications for $P_{\rm eff}=\mp1$}

As has been emphasized in the notable paper by Parke and Shadmi~\cite{ps96},
the Born term polarization formulas considerably simplify for the case of
maximal effective beam polarization $P_{\rm eff}=-1$ which corresponds to a
$(e^-_L,e^+_R)$ configuration. Although designed for the case of top--antitop
spin--spin correlations, the results of Ref.~\cite{ps96} are easily adopted to
the case of single-spin polarization as also noted in
Ref.~\cite{Kodaira:1998gt}. From a practical point of view the limiting case
$P_{\rm eff}=-1$ is very interesting since, as was emphasized in Sec.~2, one
can get quite close to the maximal value $P_{\rm eff}=-1$ even if the beam
polarizations are not close to their maximal values. Similar simplifications
occur for the case $P_{\rm eff}=+1$. In order to distinguish between the two
cases we add the suffices $LR$ and $RL$ for quantities derived for the case
$P_{\rm eff}=-1$ and $P_{\rm eff}=+1$, respectively.

For the Born term case and in the limit $P_{\rm eff}=-1$, the polarized
numerators~(\ref{lpol}) and~(\ref{trpol}) take a factorized form,
\begin{eqnarray}
\label{psell}
N_{LR}^{(\ell)}(\cos\theta)&=&-\frac38\bigg(f_{LL}(\cos\theta+v)
  +f_{LR}(\cos\theta-v)\bigg)\,A_{LR}(\cos\theta)\,\,2N_{c}s\,,\\ 
\label{pstr}
N_{LR}^{(tr)}(\cos\theta)&=&\frac38\sin\theta\sqrt{1-v^2}\,
(f_{LL}+f_{LR})\,A_{LR}(\cos\theta)\,\,2N_{c}s\,,
\end{eqnarray}
where the common factor $A_{LR}(\cos\theta)$ is given by 
\begin{equation}
\label{psA}
A_{LR}(\cos\theta)=f_{LL}(1+v\cos\theta)+f_{LR}(1-v\cos\theta)\,.
\end{equation}
We have made use of the chiral electroweak coupling coefficients $f_{LL/LR}$ of
Ref.~\cite{ps96} which are simply related to our electroweak coupling factors
$g_{ij}$ (see the Appendix). One can check that one can obtain
$N_{LR}^{(\ell)}(\cos\theta)$ in Eq.~(\ref{psell}) from the generic spin
formula Eq.~(1) of Ref.~\cite{ps96} when one specifies to the helicity system
with $\cos\xi=+1$. Similarly, one obtains $N_{LR}^{(tr)}(\cos\theta)$ in
Eq.~(\ref{pstr}) when one specifies to the transversity system $\cos\xi=0$.
In each of the two respective systems, one has to take the cross section
difference $\sigma(t\!\uparrow)-\sigma(t\!\downarrow)$.

One can then determine the angle $\alpha$ enclosing the direction of the top
quark and its polarization vector by taking the ratio $N^{(tr)}/N^{(\ell)}$.
One has
\begin{equation}
\label{offdiag}
\tan\alpha_{LR}=\frac{N_{LR}^{(tr)}(\cos\theta)}{N_{LR}^{(\ell)}(\cos\theta)}
  =-\frac{\sin\theta\sqrt{1-v^2}\,(f_{LL}+f_{LR})}{f_{LL}(\cos\theta+v)
  +f_{LR}(\cos\theta-v)}\,.
\end{equation}
For example, at threshold ($v=0$) one has $\tan\alpha_{LR}=-\tan\theta$ with
the solution $\alpha_{LR}=180^\circ-\theta$ in agreement with the corresponding
limit in Sec.~4. As another example we take $\theta=90^\circ$ and obtain
$\tan\alpha_{LR}=-(f_{LL}+f_{LR})/(f_{LL}-f_{LR})\cdot\sqrt{1-v^2}/v$. For
$\sqrt s=500\GeV$ this gives $\alpha_{LR}=124.9^\circ$, i.e.\ the polarization
vector is still close to its threshold value of $\alpha_{LR}=90^\circ$ but
has started to turn to its asymptotic value of $\alpha_{LR}=180^\circ$.

Equation~(\ref{offdiag}) is nothing but the defining equation for the
off-diagonal basis in Ref.~\cite{ps96} considering the fact that their angle
$\xi$ is related to $\alpha_{LR}$ by $\xi=180^\circ-\alpha_{LR}$. In the
coordinate system where the $z$ axis is defined by the angle $\alpha_{LR}$
given in Eq.~(\ref{offdiag}), the polarization vector of the top quark is
purely longitudinal. In particular, this means that its transverse component
is zero in the off-diagonal basis implying that the density matrix of the top
quark is diagonal in this basis. In this sense the ``off-diagonal'' basis is a
diagonal basis and the wording ``off-diagonal'' used in Ref.~\cite{ps96} for
this basis can lead to a misunderstanding.

A different but equivalent view on the off-diagonal basis may be obtained by
rotating the nondiagonal helicity system density matrix of the top quark
($m,n=\pm 1/2$)
\begin{equation}
\label{dmatrix1}
\rho_{mn} =\frac12\left(\begin{array}{cc}
  1+P^{(\ell)}&P^{(tr)}\\P^{(tr)}&1-P^{(\ell)}
  \end{array}\right)\\
  =\frac12(\sigma\cdot\bbbone+\vec{\xi}\cdot\vec{\sigma}\,).
\end{equation}
in the scattering plane by an angle $\alpha$. One has
\begin{eqnarray}
\label{dmatrix2}
\rho'_{m'n'}&=&d^{1/2}_{m'm}(\alpha)\,\rho_{mn}\,d^{1/2\dagger}_{nn'}(\alpha)
  \nonumber\\[.3cm]
  &=&\frac12\left(\begin{array}{cc}
  1+(P^{(\ell)}\cos\alpha+P^{(tr)}\sin\alpha)&
  P^{(\ell)}\sin\alpha-P^{(tr)}\cos\alpha\\ 
  P^{(\ell)}\sin\alpha-P^{(tr)}\cos\alpha&
  1-(P^{(\ell)}\cos\alpha+P^{(tr)}\sin\alpha)
  \end{array}\right)\nonumber\\[.2cm]&&
  \qquad\to\qquad\frac12\left(\begin{array}{cc}
   1+|\vec{P}\,|&0\\0&1-|\vec{P}\,|\end{array}\right)\,,
\end{eqnarray}
where $d^{1/2}_{m'm}(\alpha)$ is the usual spin-$1/2$ Wigner rotation matrix
and $|\vec{P}\,|=\sqrt{P^{(\ell)2}+P^{(tr)2}}$. It is evident that a rotation
by the angle $\alpha=\alpha_{LR}$ defined in Eq.~(\ref{offdiag}) diagonalizes
the original density matrix as indicated in the last line of
Eq.~(\ref{dmatrix2}).

\begin{figure}
\begin{center}
\epsfig{figure=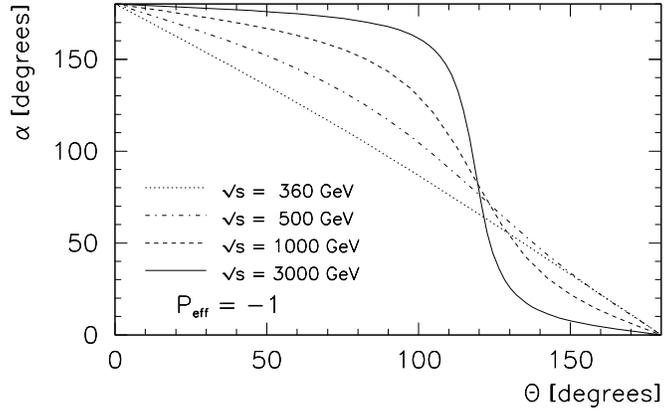, scale=0.55}\\[12pt]\qquad(a)\\[12pt]
\epsfig{figure=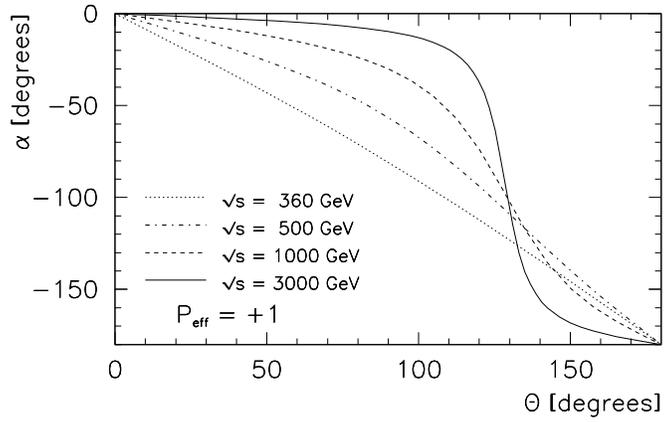, scale=0.55}\\[12pt]\qquad(b)\\[12pt]
\epsfig{figure=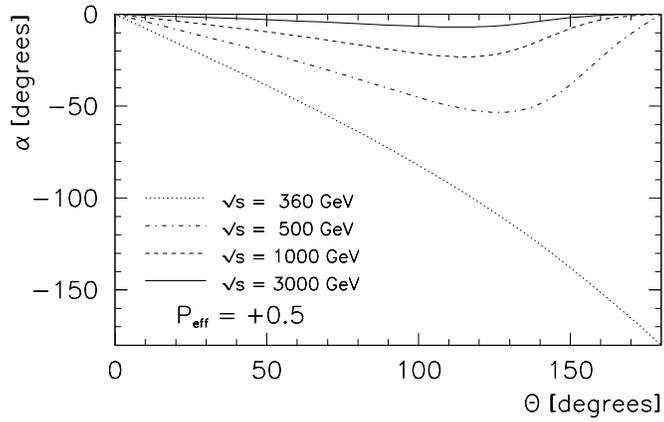, scale=0.55}\\[12pt]\qquad(c)
\end{center}
\caption{Correlation of angles $\alpha$ and $\theta$ for (a)
  $P_{\rm eff}=-1$ ($\alpha=\alpha_{LR}$), (b) $P_{\rm eff}=+1$
  ($\alpha=\alpha_{RL}$), and (c) $P_{\rm eff}=+0.5$ for different values
  of the c.m.\ energy $\sqrt s=360$, $500$, $1000$, and $3000\GeV$ (notation
  as in Fig.~\ref{sigmar})
\label{alphath}}
\end{figure}

The correlation between the angles $\alpha$ (for general values of
$P_{\rm eff}$) and $\theta$ are shown in the contour plots Fig.~\ref{alphath}.
In Fig.~\ref{alphath}a we choose $P_{\rm eff}=-1$ and show fixed energy
contours in the $(\alpha,\theta)$-plane for several values of the c.m.\ energy
$\sqrt s$. Up to $\sqrt s=1000\GeV$ the correlations do not deviate very much
from the threshold correlation $\alpha=180^\circ-\theta$. In the limit $v=1$
one has a steplike behaviour of the correlation function as discussed before
in Sec.~4. In Fig.~\ref{alphath}b we show the same plots for $P_{\rm eff}=+1$.
The approach of the correlation curves to the steplike behaviour at $v=1$ is
somewhat faster than in the case $P_{\rm eff}=-1$. In Fig.~\ref{alphath}c
we show the same curves for $P_{\rm eff}=0.5$ where one is close to the
polarization zero. The $\sqrt s=360\GeV$ correlation curve is still close
to the corresponding threshold curve $\alpha=-\theta$. At higher energies one 
sees a different behaviour in as much as the correlation curves run into
$\alpha=0^{\circ}$ at the backward point as does the flat asymptotic curve as
discussed before in Sec.~4 (Fig.~\ref{f:fig7}).

In order to calculate the normalized polarization components one needs also
the denominator factor $D(\cos\theta)$ in Eq.~(\ref{D}), again for the Born
term case and $P_{\rm eff}=-1$. One has
\begin{equation}
\label{psD}
D_{LR}(\cos\theta)=\frac38\Big(A_{LR}^2-2f_{LL}f_{LR}\,v^2\sin^2\theta\Big)\,
\,2N_{c}s\,.
\end{equation}
proportional to the cross section sum
$\sigma(t\!\uparrow)+\sigma(t\!\downarrow)$ in any of the systems in
Ref.~\cite{ps96}.

Using Eqs.~(\ref{psell}) and~(\ref{psD}) the longitudinal polarization
$P^{(\ell)}_{LR}=N^{(\ell)}_{LR}/D_{LR}$ can be seen to become maximally $-1$
and $+1$ in the forward and backward directions, respectively, in agreement
with angular momentum conservation as before. One also reproduces the
threshold formula Eq.~(\ref{thresh}) and the high-energy formula
Eq.~(\ref{helimit}) when these are specified to $P_{\rm eff}=-1$. The
longitudinal polarization goes through zero at
\begin{equation}
\label{pscos}
\cos\theta_0=-\frac{f_{LL}-f_{LR}}{f_{LL}+f_{LR}}\,v
  =\frac{g_{14}-g_{44}}{g_{11}-g_{41}+g_{12}-g_{42}}\,v\quad(\,=-0.48v)\,.
\end{equation}
At this value of $\cos\theta$ the polarization vector of the top quark is
orthogonal to its momentum. Later on we shall see that, at this point,
$P^{(tr)}$ acquires its maximal value and $|\vec{P}\,|$ acquires its minimal
value. Since the ratio $(f_{LL}-f_{LR})/(f_{LL}+f_{LR})$ is only mildly
energy-dependent, the location of the zero is mainly determined by the velocity
of the top quark, i.e.\ it moves towards the backward point when the energy is
increased. For convenience we have added the $\sqrt s=500\GeV$ value of the
electroweak coupling ratio in brackets in Eq.~(\ref{pscos}).
 
The transverse polarization $P^{(tr)}_{LR}$ vanishes in the forward and
backward directions due to angular momentum conservation, as is explicit in
Eq.~(\ref{pstr}). It becomes maximal at the point where the longitudinal
polarization goes through zero. This can be verified by an explicit
calculation, {\it viz.}
\begin{equation}
\frac{d\,P^{(tr)}_{LR}}{d\,\cos\theta}\,\Bigg|_{\displaystyle\,\cos\theta_0}
\,=\,\,0\,,
\end{equation}
where $\cos\theta_0$ is given in Eq.~(\ref{pscos}).

Whereas there are no illuminating expressions for the longitudinal and
transverse polarization components for general values of the velocity $v$, the
magnitude of the polarization $|\vec{P}\,|$ for $P_{\rm eff}=-1$ takes the
simple form
\begin{equation}
\label{expanda}
|\vec{P}_{LR}|=\frac{\sqrt{N_{LR}^{(\ell)2}+N_{LR}^{(tr)2}}}{D_{LR}}
  =\frac{\sqrt{1-4a_{LR}}}{1-2a_{LR}}=1-2a_{LR}^2-8a_{LR}^3-18a_{LR}^3\ldots,
\end{equation}
where the coefficient $a_{LR}$ depends on $\cos\theta$ through
\begin{equation}
a_{LR}(\cos\theta)=\frac{f_{LL}f_{LR}}{A_{LR}^2(\cos\theta)}v^2\sin^2\theta\,.
\end{equation}
The convergence of the expansion in Eq.~(\ref{expanda}) is rather slow except
for very small values of $a_{LR}$. Note that the expansion in
Eq.~(\ref{expanda}) deviates from $1$ only at $O(a_{LR}^2)$. At the forward
and backward points where $a_{LR}=0$, one has $|\vec{P}_{LR}|=1$ as stated
before. Between the forward and backward points the polarization remains
reasonably large. For example, for $\sqrt s=500\GeV$ the polarization never
drops below $|\vec{P}_{LR}|=0.95$. Differentiating of Eq.~(\ref{expanda}) with
respect to $\cos\theta$ one can see that the minimum of $|\vec{P}_{LR}|$ occurs
at the point where the longitudinal polarization $P^{(\ell)}_{LR}$ vanishes
(see Eq.~(\ref{pscos})), i.e.\ where the polarization is purely transverse.
The high-energy limit of Eq.~(\ref{expanda}) is discussed in Sec.~4.

Similar simplifications occur for the case $P_{\rm eff}=+1$ which corresponds
to the $(e^-_R,e^+_L)$ configuration. This is effected by the replacement
$f_{LL}\to f_{RR}$ and $f_{LR}\to f_{RL}$ with a corresponding change in the
notation $A_{LR},a_{LR} \to A_{RL},a_{RL}$. Further one has
$N_{RL}^{(\ell)}=-N_{LR}^{(\ell)}(f_{LL}\to f_{RR}; f_{LR}\to f_{RL})$ and
$N_{RR}^{(tr)}=-N_{LL}^{(tr)}(f_{LL}\to f_{RR}; f_{LR}\to f_{RL})$. The zero
of $P^{(\ell)}$ is now located at $\cos\theta_0=-0.63v$ for $\sqrt s=500\GeV$,
i.e.\ the zero is closer to the backward point than in the case
$P_{\rm eff}=-1$. For $\theta=90^\circ$ the angle $\alpha_{RL}$ can be
calculated from
$\tan\alpha_{RL}=-(f_{RR}+f_{RL})/(f_{RR}-f_{RL})\cdot\sqrt{1-v^2}/v$ which, at
$\sqrt s=500\GeV$, gives $\alpha_{RL}=-48^\circ$. At $\cos\theta=0$ and
$\sqrt s=500\GeV$, one has $a_{LR}>a_{RL}$ leading to
$|\vec{P}_{LR}|<|\vec{P}_{RL}|$, i.e.\ the $(e^-_R,e^+_L)$ configuration leads
to larger values of the polarization than the $(e^-_L,e^+_R)$ configuration at
this point of parameter space. In Fig.~\ref{alphath}b we show a contour plot
in the $(\alpha_{RL},\theta)$ plane for several values of the c.m.\ energy
$\sqrt s$. As is the case for the $(\alpha_{LR},\theta)$ correlations, the
$(\alpha_{RL},\theta)$ correlations do not deviate very much from the
threshold correlation $\alpha=-\theta$ up to $\sqrt s=1000\GeV$.

\begin{figure}[t]
\begin{center}
\epsfig{figure=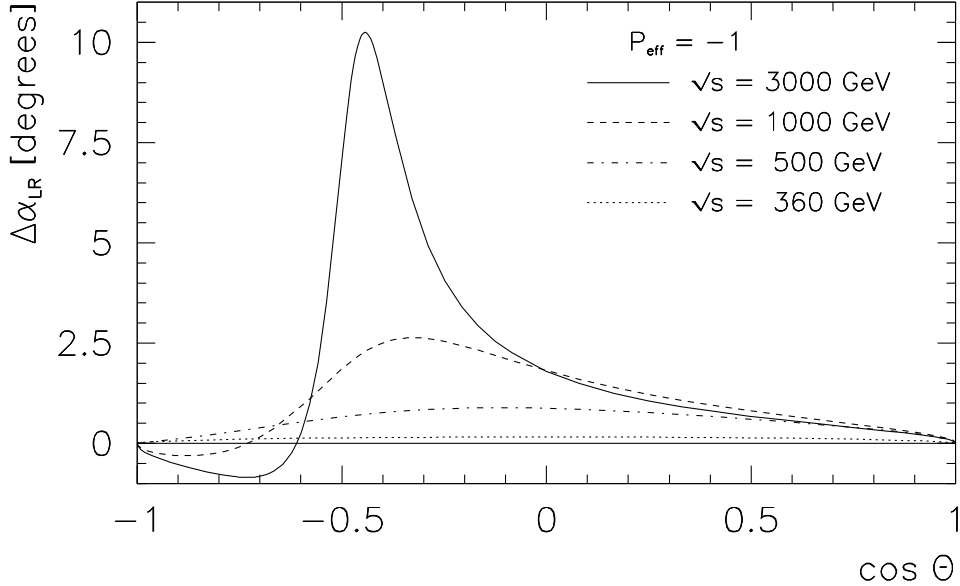, scale=0.8}
\end{center}
\caption{Difference
  $\Delta\alpha_{LR}=\alpha_{LR}({\rm NLO})-\alpha_{LR}({\rm LO})$ of NLO and
  LO polarization angles for $\sqrt s=360$, $500$, $1000$, and $3000\GeV$
  (notation as in Fig.~\ref{sigmar})
 \label{f:alphat1m}}
\end{figure}

As a last point we discuss how the polarization angle $\alpha_{LR}$ changes
when going from LO to NLO. In Fig.~\ref{f:alphat1m} we show a plot of the
$\cos\theta$ dependence of the difference
$\Delta\alpha_{LR}=\alpha_{LR}({\rm NLO})-\alpha_{LR}({\rm LO})$ for 
different energies. The maximal values of the difference occur at values of
$\cos\theta$ where the polarization vector is perpendicular to the top quark's
momentum, i.e.\ where $\alpha_{LR}=90^\circ$ (see discussion after
Eq.~(\ref{pscos})). The difference can become as big as $10^\circ$ for
$\sqrt s=3000\GeV$. The radiative corrections can thus be seen to rotate the
polarization vector away from the off-diagonal basis by a nonnegligible amount.

\section{The polarization components $P^{(\ell)}$, $P^{(tr)}$, and $P^{(n)}$}

We now turn to the numerical discussion of the three polarization components
$P^{(\ell)}$, $P^{(tr)}$, and $P^{(n)}$ keeping in mind the Born term
simplifications for $P^{(\ell)}$ and $P^{(tr)}$ discussed in Secs.~4 and 5. We
start our discussion with the longitudinal component $P^{(\ell)}$.
\begin{figure}
\begin{center}
\epsfig{figure=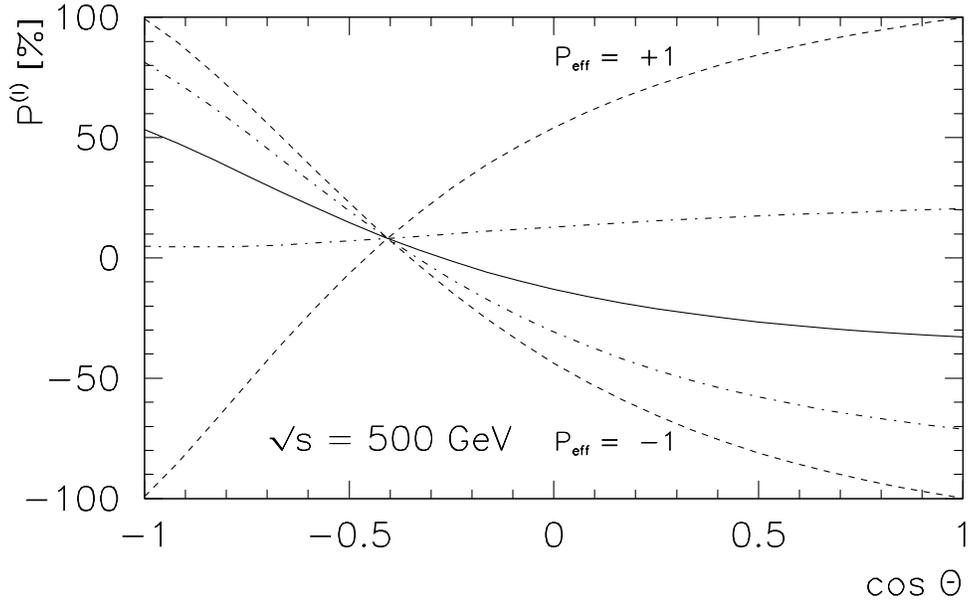, scale=0.8}\\[12pt]\qquad(a)\\[12pt]
\epsfig{figure=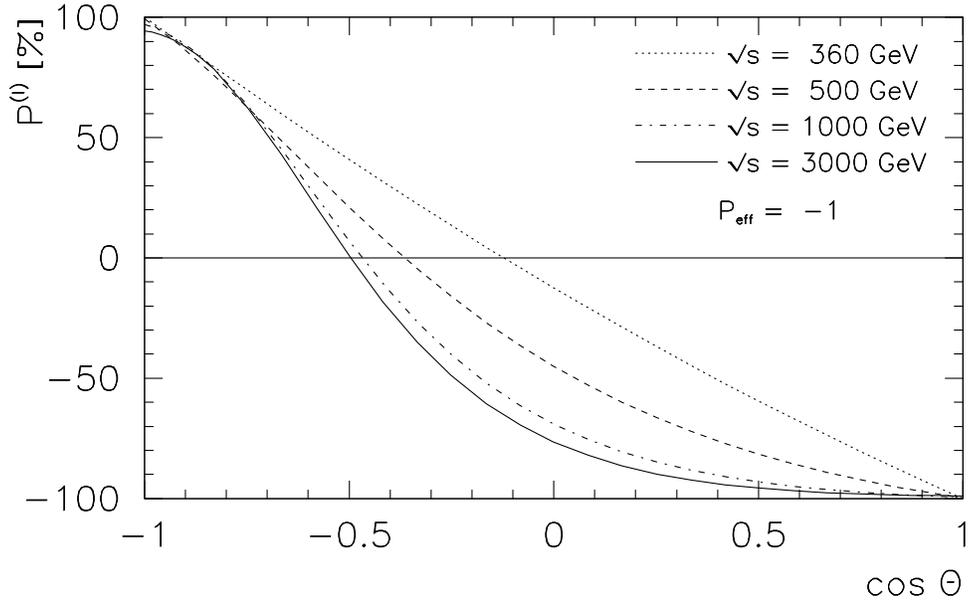, scale=0.8}\\[12pt]\qquad(b)
\end{center}
\caption{NLO longitudinal top polarization as a function of $\cos\theta$ drawn
  (a) for effective beam polarizations $P_{\rm eff}=-1,-0.5,0,+0.5,+1$
  (notation as in Fig.~\ref{f:fig1}) at $\sqrt s=500\GeV$; (b) at beam
  energies $\sqrt s=360$, $500$, $1000$, and $3000\GeV$ (notation as in
  Fig.~\ref{sigmar}) and $P_{\rm eff}=-1$
\label{poll}}
\end{figure}
In Fig.~\ref{poll}a we show the dependence of the NLO longitudinal
polarization $P^{(\ell)}$ on $\cos\theta$ at $\sqrt s=500\GeV$ for several
values of $P_{\rm eff}$ spanning the whole parameter range of $P_{\rm eff}$.
The dependence of $P^{(\ell)}$ on $P_{\rm eff}$ and $\cos\theta$ is quite
pronounced. For $P_{\rm eff}=\pm1$ the $\cos\theta$-dependence already
deviates considerably from the (Born term) threshold behaviour
$P^{(\ell)}=\pm\cos\theta$. It is quite interesting to observe that all NLO
curves intersect at one point where $\cos\theta=-0.406$. This can be verified
by setting to zero the derivative of $P^{(\ell)}$ with respect to
$P_{\rm eff}$. The relevant higher order equation admits of a solution at the
above value of $\cos\theta$. In Fig.~\ref{poll}b we show the
$\cos\theta$-dependence of $P^{(\ell)}$ for several energies keeping
$P_{\rm eff}$ fixed at $P_{\rm eff}=-1$. At the resolution of the figure all
curves seemingly go through $-1$ and $+1$ at the forward and backward point,
respectively, showing that hard gluon emission effects are not very strong at
these energies. The energy dependence is not very pronounced, even if the
$\sqrt s=500\GeV$ curve already deviates from the threshold behaviour
$P^{(\ell)}=-\cos\theta$.

The strong dependence of $P^{(\ell)}$ on $P_{\rm eff}$ can be nicely exposed
by considering the LO expression for the polar mean
$\langle P^{(\ell)}\rangle$ which is obtained by integrating the numerator and
the denominator in Eq.~(\ref{NoverD}) separately over $\cos\theta$. One obtains
\begin{equation}
\langle P^{(\ell)}\rangle=\frac{\langle N^{(\ell)}\rangle}{\langle D\rangle}
  =\frac43v\frac{g_{14}+g_{44}P_{\rm eff}}{(g_{11}+g_{41}P_{\rm eff})(1+v^2/3)
  +(g_{12}+g_{42}P_{\rm eff})(1-v^2)}\,.
\end{equation}
$\langle P^{(\ell)}\rangle$ vanishes at threshold. In the high-energy limit,
one has $\langle\! P^{(\ell)}\rangle\!=\!(g_{14}+g_{44}P_{\rm eff})/
(g_{11}+\!g_{41}P_{\rm eff})$ which, for $P_{\rm eff}=\pm1$, gives
$\langle P^{(\ell)}\rangle=0.882$ and $\langle P^{(\ell)}\rangle=-0.766$ close
to the $\sqrt s=1000\GeV$ values in Fig.~\ref{f:fig1}. At $\sqrt s=500\GeV$,
one has
\begin{equation}
\langle P^{(\ell)}\rangle=\langle P^{(\ell)}\rangle(P_{\rm eff}=0)\,\,
\frac{1-3.61\,P_{\rm eff}}{1-0.37\,P_{\rm eff}}
  =\left\{\begin{array}{rl}
  +0.62&\quad P_{\rm eff}=+1\\
  -0.15&\quad\phantom{P_{\rm eff}}=\phantom{+}0\\
  -0.50&\quad\phantom{P_{\rm eff}}=-1
  \end{array}\right\}\,.
\end{equation}
One observes a strong dependence of the mean longitudinal polarization on
$P_{\rm eff}$. By comparing with the $\sqrt s=500\GeV$ point in
Fig.~\ref{f:fig1}a, one observes a $2\%$ change in $\langle P^{(\ell)}\rangle$
due to the radiative corrections.

\begin{figure}
\begin{center}
\epsfig{figure=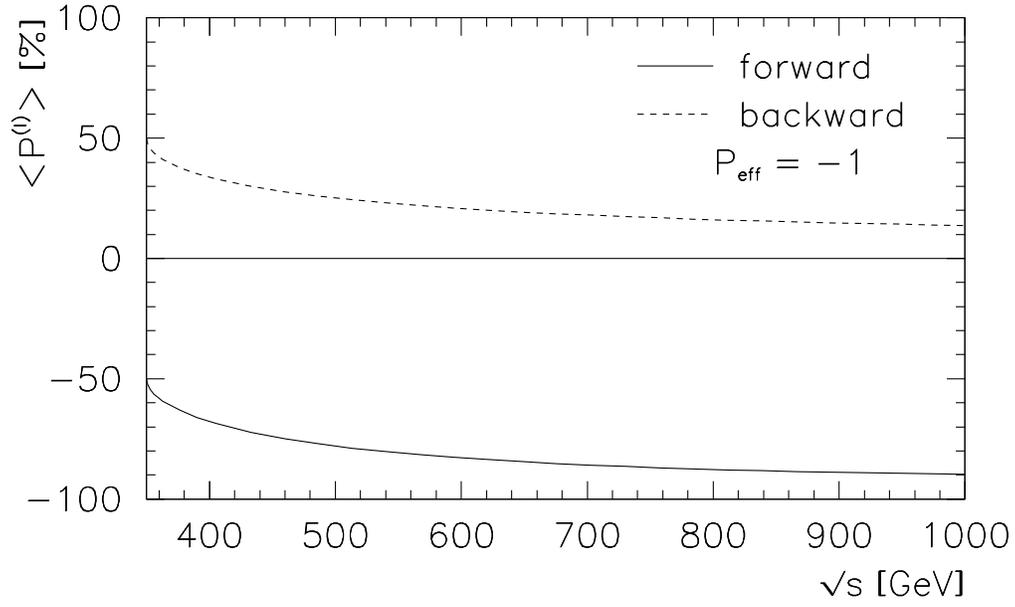, scale=0.8}\\[12pt]\qquad(a)\\[12pt]
\epsfig{figure=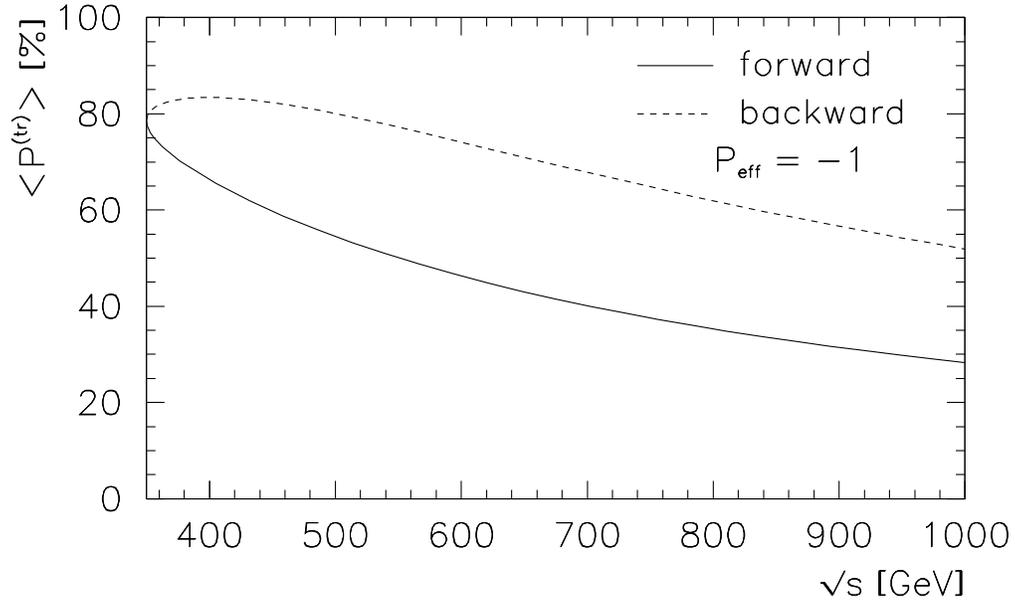, scale=0.8}\\[12pt]\qquad(b)
\end{center}
\caption{Average (a) longitudinal polarization $\langle P^{(\ell)}\rangle$
  and (b) transverse polarization $\langle P^{(tr)}\rangle$ in the
  forward and backward hemispheres for $P_{\rm eff}=-1$
  \label{f:fig12}}
\end{figure}

The same strong dependence on $P_{\rm eff}$ is found when one averages over
the forward hemisphere where one has ($\sqrt s=500\GeV$)
\begin{equation}
\label{Faverage}
\langle P^{(\ell)}\rangle_F=\langle P^{(\ell)}\rangle_F\,\,
  (P_{\rm eff}=0)\,\,\frac{1-3.09\,P_{\rm eff}}{1-0.34\,P_{\rm eff}}
  =\left\{\begin{array}{rl}
  +0.85&\quad P_{\rm eff}=+1\\
  -0.27&\quad\phantom{P_{\rm eff}}=\phantom{+}0\\
  -0.81&\quad\phantom{P_{\rm eff}}=-1
  \end{array}\right\}\,.
\end{equation}
When one averages over the backward hemisphere the average longitudinal
polarization is smaller and the dependence on $P_{\rm eff}$ is much weaker
{\it viz.}
\begin{equation}
\label{Baverage}
\langle P^{(\ell)}\rangle_B=\langle P^{(\ell)}\rangle_B\,\,
  (P_{\rm eff}=0)\,\,\frac{1-1.04\,P_{\rm eff}}{1-0.43\,P_{\rm eff}}
  =\left\{\begin{array}{rl}
  -0.01&\quad P_{\rm eff}=+1\\
  +0.13&\quad\phantom{P_{\rm eff}}=\phantom{+}0\\
  +0.18&\quad\phantom{P_{\rm eff}}=-1
  \end{array}\right\}\,.
\end{equation}

In Fig.~\ref{f:fig12}a we show a plot of the energy dependence of the forward
and backward averages of the longitudinal polarizations for $P_{\rm eff}=-1$.
The forward average $\langle P^{(\ell)}\rangle_F$ is large and negative. It
starts with a nominal threshold value of $\langle P^{(\ell)}\rangle_F=-0.5$
and slowly drops to a value of $\langle P^{(\ell)}\rangle_F=-0.90$ at
$\sqrt s=1000\GeV$ which is not far from the asymptotic Born term value
$\langle P^{(\ell)}\rangle_F=-(7f_{LL}^2-f_{LR}^2)/(7f_{LL}^2+f_{LR}^2)=-0.96$.
The backward average $\langle P^{(\ell)}\rangle_B$ is smaller and positive. It
drops from a nominal threshold value of $\langle P^{(\ell)}\rangle_B=+0.5$ to
$\langle P^{(\ell)}\rangle_B=+0.14$ at $\sqrt s=1000\GeV$ as compared to the
asymptotic Born term value
$\langle P^{(\ell)}\rangle_B=-(f_{LL}^2-7f_{LR}^2)/(f_{LL}^2+7f_{LR}^2)=-0.04$.

\begin{figure}
\begin{center}
\epsfig{figure=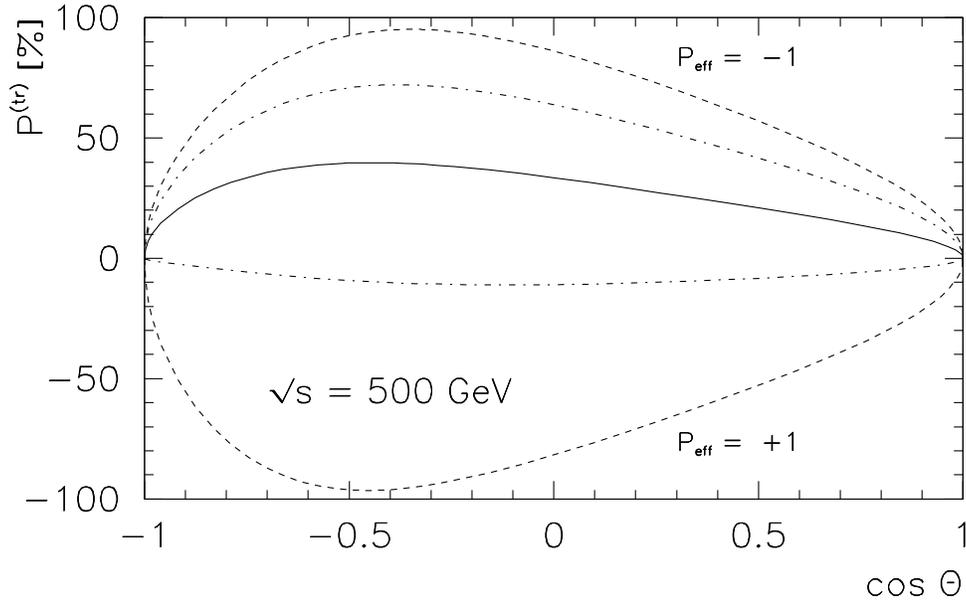, scale=0.8}\\[12pt]\qquad(a)\\[12pt]
\epsfig{figure=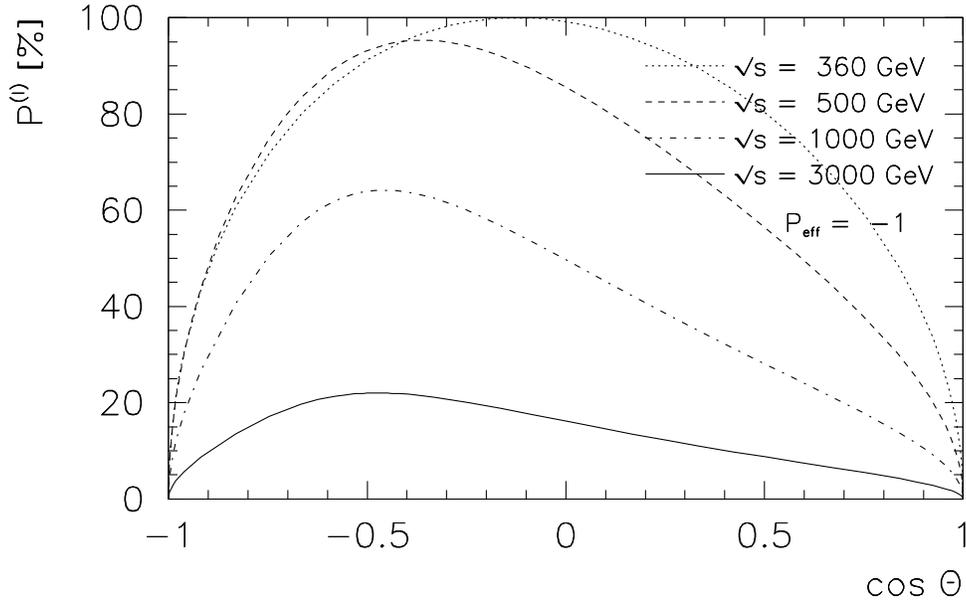, scale=0.8}\\[12pt]\qquad(b)
\end{center}
\caption{NLO transverse top polarization as a function of the scattering angle
  $\theta$ drawn (a) for effective beam polarizations
  $P_{\rm eff}=-1,-0.5,0,+0.5,+1$ (notation as in Fig.~\ref{f:fig1}) at
  $\sqrt s=500\GeV$; (b) at beam energies $\sqrt s=360$, $500$, $1000$,
  and $3000\GeV$ (notation as in Fig.~\ref{sigmar}) and $P_{\rm eff}=-1$
  \label{f:fig6}}
\end{figure}

We now turn to the transverse polarization component $P^{(tr)}$. Similar to
Figs.~\ref{poll}a and~\ref{poll}b we show the corresponding curves for
$P^{(tr)}$ in Figs.~\ref{f:fig6}a and~\ref{f:fig6}b. The transverse
polarization vanishes at the end points due to the overall $\sin\theta$ factor
in the angular decay distribution Eq.~(\ref{trpol}). The dependence on
$P_{\rm eff}$ is again quite pronounced. One observes a faster change with
$P_{\rm eff}$ at $P_{\rm eff}=+1$ than at $P_{\rm eff}=-1$. $P^{(tr)}$
vanishes close to  $P_{\rm eff}=+0.5$. At $\sqrt s=500\GeV$ the deviations
from the threshold behaviour $P^{(tr)}=\pm\sqrt{1-\cos\theta^2}$ for
$P_{\rm eff}=\pm1$ are slight but clearly visible. The $(\sqrt s)^{-1}$
dependence of the transverse polarization is easily discernible in
Fig.~\ref{f:fig6}b. A visual inspection shows that, as is exact in the Born
term case for $P_{\rm eff}=\pm1$, the NLO longitudinal and transverse
polarization components are complementary in the sense that the transverse
polarization becomes maximal very close to the point where the longitudinal
polarization becomes minimal, and vice versa. This observation bodes well for
the existence of large values of the total polarization as discussed in Sec.~7.

For the mean value of the transverse polarization, one obtains the Born term
level expression
\begin{equation}
\langle P^{(tr)}\rangle=-\frac\pi2\frac{m_t}{\sqrt s}
\frac{g_{41}+g_{42}+(g_{11}+g_{12})P_{\rm eff}}{(g_{11}+g_{41}P_{\rm eff})
  (1+v^2/3)+(g_{12}+g_{42}P_{\rm eff})(1-v^2)}\,.
\end{equation}
At nominal threshold one has $\langle P^{(tr)}\rangle=\mp \pi/4$ for
$P_{\rm eff}=\pm1$ close to what is seen in Fig.~\ref{f:fig1}. In the high
energy limit $\langle P^{(tr)}\rangle$ vanishes as $(\sqrt s)\,^{-1}$.
At $\sqrt s=500\GeV$ one has
\begin{equation}
\langle P^{(tr)}\rangle=\langle P^{(tr)}\rangle(P_{\rm eff}=0)\,\,
\frac{1-2.53\,P_{\rm eff}}{1-0.37\,P_{\rm eff}}
  =\left\{\begin{array}{rl}
  -0.57&\quad P_{\rm eff}=+1\\
  +0.24&\quad\phantom{P_{\rm eff}}=\phantom{+}0\\
  +0.61&\quad\phantom{P_{\rm eff}}=-1
  \end{array}\right\}\,,
\end{equation}
showing again the large effect of beam polarization. By comparing with the
$\sqrt s=500\GeV$ point in Fig.~\ref{f:fig1}b, one observes a $1.5\%$ change
in $\langle P^{(tr)}\rangle$ due to the radiative corrections.

For the ratio of the forward and backward mean of the transverse polarization,
one obtains 
\begin{equation}
\label{FBtr}
\frac{\langle P^{(tr)}\rangle_F}{\langle P^{(tr)}\rangle_B}
  =\frac{\langle N^{(tr)}\rangle_F}{\langle N^{(tr)}\rangle_B}
  \,\,\frac{\langle\sigma\rangle_B}{\langle\sigma\rangle_F}
  =\left\{\begin{array}{rl}
  +0.54&\quad P_{\rm eff}=+1\\
  +0.53&\quad\phantom{P_{\rm eff}}=\phantom{+}0\\
  +0.61&\quad\phantom{P_{\rm eff}}=-1
  \end{array}\right\}\,,
\end{equation}
There is a slight dominance of the backward mean as also evident in
Fig.~\ref{f:fig6}. The dependence of the ratio~(\ref{FBtr}) on $P_{\rm eff}$
is not very pronounced. 

A plot of the energy dependence of the forward and backward averages of the
transverse polarization is shown in Fig.~\ref{f:fig12}b. Both curves start at
the nominal threshold value
$\langle P^{(tr)}\rangle_F=\langle P^{(tr)}\rangle_B=\pi/4$ and then quite
slowly begin their descent to their asymptotic demise. At $\sqrt s=500\GeV$
and $P_{\rm eff}=-1$, one can compare the NLO result for
$\langle P^{(tr)}\rangle_F/\langle P^{(tr)}\rangle_B=0.68$ with the
corresponding LO result
$\langle P^{(tr)}\rangle_F/\langle P^{(tr)}\rangle_B=0.61$ in Eq.~(\ref{FBtr}).

The normal polarization component $P^{(n)}$ is a $T$-odd observable and thus
obtains only contributions from the imaginary parts of the production
amplitudes. Since we are neglecting the contribution from the imaginary part
of the $Z$ propagator the only contribution to the normal polarization
component $P^{(n)}$ at $O(\alpha_{s})$ is that of the imaginary part of the
one-loop contributions (Eqs.~(\ref{loopI}) and~(\ref{loopA})).
When averaging over $\cos\theta$, the contributions of
$H_I^{1,2(n)}({\it loop})$ drop out and one has the $O(\alpha_s)$ result
\begin{equation}
\langle P^{(n)}\rangle=-\alpha_s\frac\pi6\frac{m_t}{\sqrt s}\,(2-v^2)
  \frac{g_{44}+g_{14}P_{\rm eff}}{(g_{11}+g_{41}P_{\rm eff})(1+v^2/3)
  +(g_{12}+g_{42}P_{\rm eff})(1-v^2)}\,.
\end{equation}
Numerically one has ($\sqrt s=500\GeV$; $\alpha_s=0.094$) 
\begin{equation}
\label{normal1}
\langle P^{(n)}\rangle=\langle P^{(n)}\rangle(P_{\rm eff}=0)\,\,
  \frac{1-0.27\,P_{\rm eff}}{1-0.37\,P_{\rm eff}}
  =\left\{\begin{array}{rl}
  -0.015&\quad P_{\rm eff}=+1\\
  -0.013&\quad\phantom{P_{\rm eff}}=\phantom{+}0\\
  -0.012&\quad\phantom{P_{\rm eff}}=-1
  \end{array}\right\}\,.
\end{equation}
Clearly the normal polarization component is small being an $O(\alpha_s)$
effect. Also, the dependence of $\langle P^{(n)}\rangle$ on the beam
polarization is quite small.

\begin{figure}
\begin{center}
\epsfig{figure=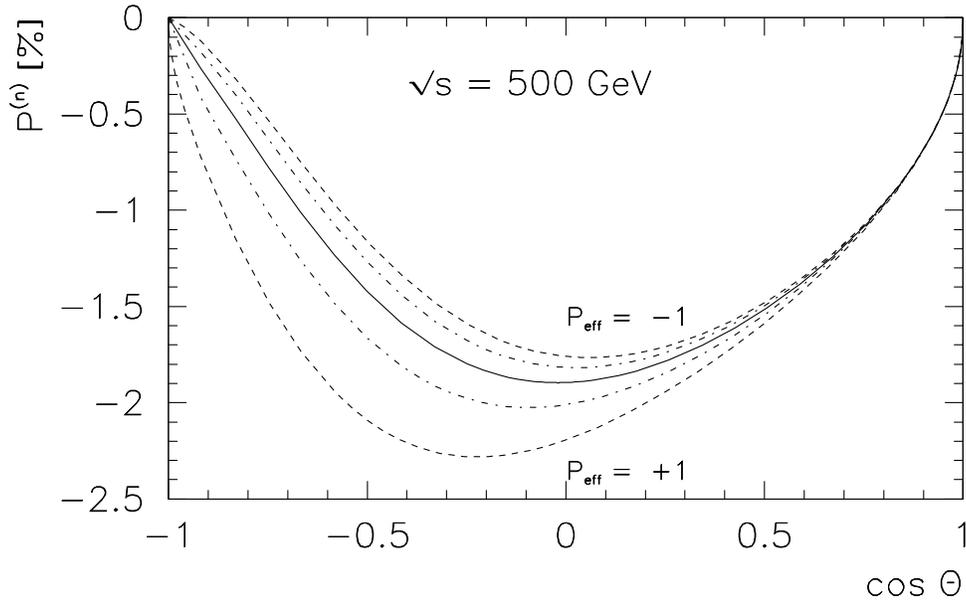, scale=0.8}\\[12pt]\qquad(a)\\[12pt]
\epsfig{figure=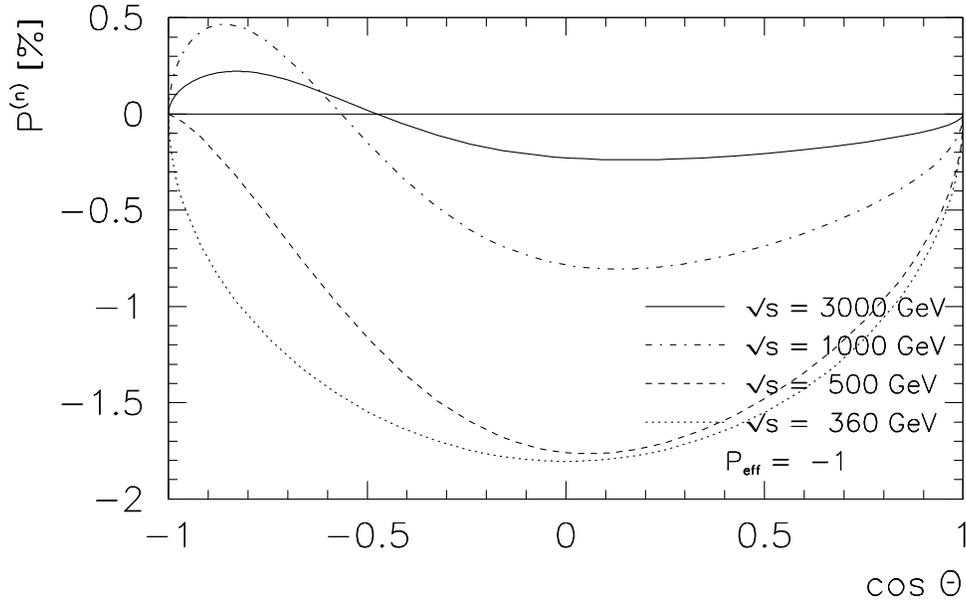, scale=0.8}\\[12pt]\qquad(b)
\end{center}
\caption{$O(\alpha_s)$ normal polarization $P^{(n)}$ of the top quark as a
  function of $\cos\theta$ (a) for effective beam polarizations
  $P_{\rm eff}=-1,-0.5,0,+0.5,+1$ (notation as in Fig.~\ref{f:fig1}) at
  $\sqrt s=500\GeV$; (b) for $P_{\rm eff}=-1$ at beam energies
  $\sqrt s=360$, $500$, $1000$, and $\sqrt s=3000\GeV$ (notation as in
  Fig.~\ref{sigmar})
  \label{poln}}
\end{figure}

In Fig.~\ref{poln} we show the $\cos\theta$ dependence of the normal component
of the polarization of the top quark. In Fig.~\ref{poln}a we keep the energy
fixed at $\sqrt s=500\GeV$ and vary $P_{\rm eff}$. The differential
distribution peaks at around $\cos\theta=0$ where the peak moves to the left
with increasing values of $P_{\rm eff}$. The peak values of $P^{(n)}$ are
around $-2\%$. The dependence on $P_{\rm eff}$ is weak as also evident in
Eq.~(\ref{normal1}). In Fig.~\ref{poln}b we plot the $\cos\theta$ dependence
for different energies keeping $P_{\rm eff}$ fixed at $P_{\rm eff}=-1$. As
expected, the normal polarization can be seen to decrease with the typical
$(\sqrt s)^{-1}$ behaviour. We mention that we are now in agreement with the
results of Ref.~\cite{rvn00} when one takes account of the fact that their
normal direction is defined opposite to ours.

Let us close this section by comparing our results to those of the authors of
Ref.~\cite{Kodaira:1998gt} who calculated $O(\alpha_s)$ radiative corrections
to rates into definite spin states in generic coordinate systems starting from
the initial beam configurations $(e^-_L,e^+_R)$ and $(e^-_R,e^+_L)$ which
correspond to $P_{\rm eff}=-1$ and $P_{\rm eff}=+1$, respectively. Put in a
different language, they compute radiative corrections to the (unnormalized)
diagonal spin density matrix elements $\sigma_{\uparrow}$ and
$\sigma_{\downarrow}$. In the helicity system, where they use the notation
$\sigma_{\downarrow/\uparrow}\equiv\sigma_{L/R}$, their polarized rate
$\sigma_{L/R}$ are related to our longitudinal polarization component
$P^{(\ell)}$ via 
\begin{equation}
\sigma_{L/R}=\frac{\sigma}2(1\mp P^{(\ell)})
\end{equation} 
Comparing to the $(e^-_L,e^+_R)$ subdominant spin rate ratio at
$\sqrt s=400\GeV$ in Table~3 of Ref.~\cite{Kodaira:1998gt} we find a $-1.34\%$
reduction relative to the LO rate ratio {\it vs.} their reduction of $-1.19\%$.
We consider the two results to be consistent with each other within rounding
errors. We mention that our NLO results have been checked before in
Ref.~\cite{rvn00}.

\begin{figure}
\begin{center}
\epsfig{figure=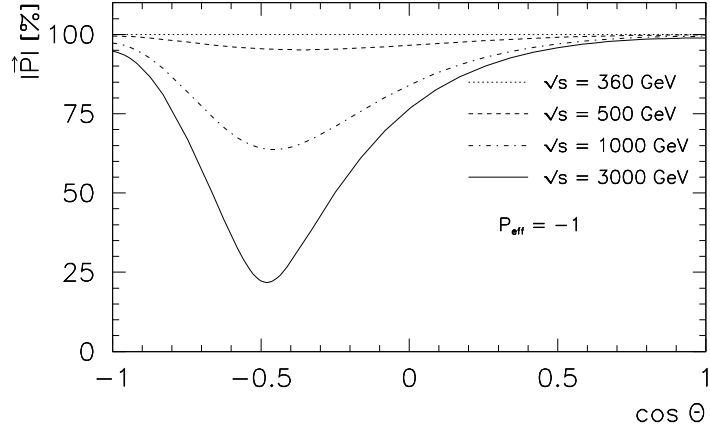, scale=0.58}\\[12pt]\qquad(a)\\[12pt]
\epsfig{figure=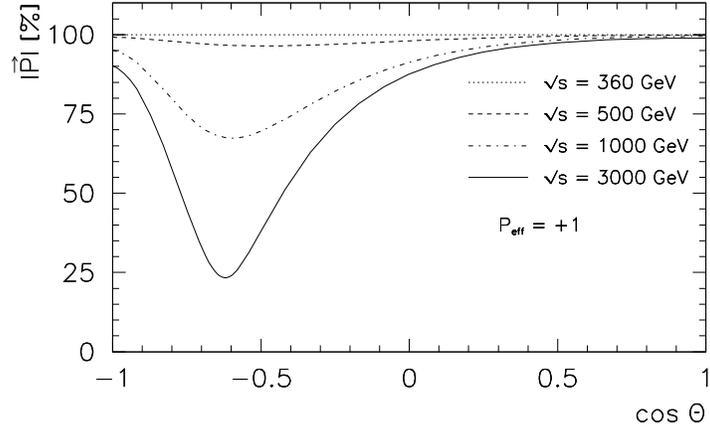, scale=0.58}\\[12pt]\qquad(b)\\[12pt]
\epsfig{figure=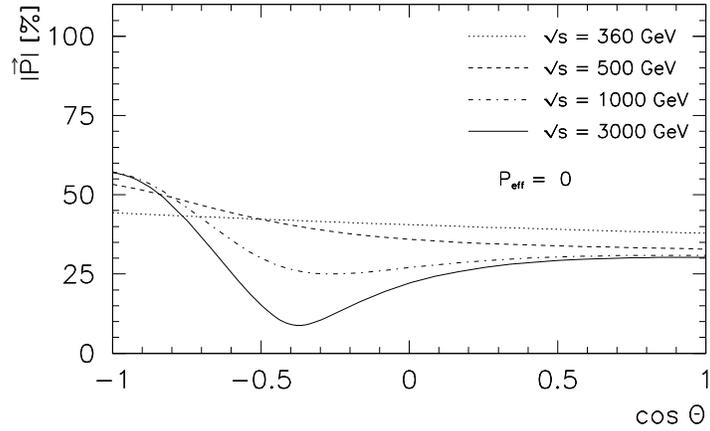, scale=0.58}\\[12pt]\qquad(c)
\end{center}
\caption{Total NLO top quark polarization as a function of $\cos\theta$ for
  beam energies $\sqrt s=360$, $500$, $1000$, and $3000\GeV$ (notation as in
  Fig.~\ref{sigmar}) and (a) $P_{\rm eff}=-1$, (b) $P_{\rm eff}=+1$,
  and (c) $P_{\rm eff}=0$
  \label{polah}}
\end{figure}

\begin{figure}[t]
\begin{center}
\epsfig{figure=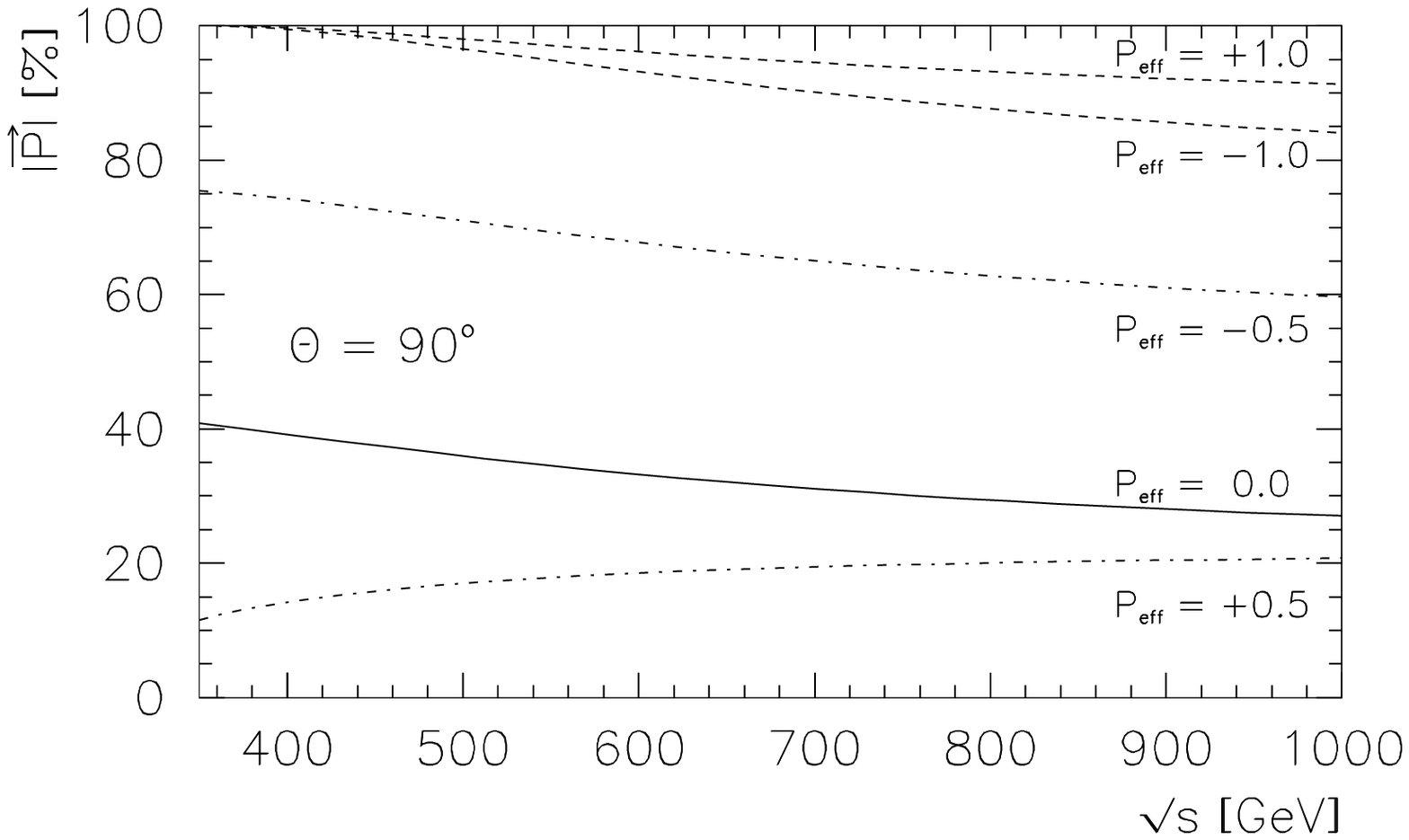, scale=0.8}
\caption{Average NLO top quark polarization $\langle|\vec{P}\,|\rangle$ for
  a scattering angle of $\theta=90^\circ$ as a function of the beam energy
  $\sqrt s$ for $P_{\rm eff}=-1,-0.5,0,+0.5,+1$ (notation as in
  Fig.~\ref{f:fig1}) 
  \label{f:fig16}}
\end{center}
\end{figure}

\section{Total polarization and\\ orientation of the polarization vector}

The magnitude of the polarization (also called total polarization) is given by
\begin{equation}
\label{totpol}
|\vec{P}\,|=\sqrt{(P^{(\ell)})^2+(P^{(tr)})^2+(P^{(n)})^2}\,.
\end{equation}
In Fig.~\ref{polah} we show the NLO dependence of $|\vec{P}\,|$ on
$\cos\theta$ at different values of $\sqrt s$ for the three different values of
$P_{\rm eff}=\pm 1$ and $0$. As a general feature one observes that the
magnitude of the polarization decreases with energy. When $P_{\rm eff}=\pm 1$
one obtains large values of $|\vec{P}\,|$, in particular in the forward
hemisphere. For example, for $\sqrt s=500\GeV$ $|\vec{P}\,|$ remains above
$95\%$ over the whole angular range for $P_{\rm eff}=\pm 1$. The polarization
is slightly larger for $P_{\rm eff}=+1$ than for $P_{\rm eff}=-1$. At
$\sqrt s=360\GeV$ and $P_{\rm eff}=\pm 1$ one is still very close to the flat
threshold behaviour $|\vec{P}\,|=1$, whereas for $P_{\rm eff}=0$ one observes
a slight falloff behaviour going from the backward to the forward point. Even
for the largest energy $\sqrt s=3000\GeV$, one does not have a zero for
$|\vec{P}\,|$ showing that one is still away from the asymptotic $v=1$ case
since asymptotically one has polarization zeros for the three cases
$P_{\rm eff}=\pm1,0$ as discussed in Sec.~4 and exhibited in Fig.~\ref{f:fig7}.
As mentioned before there is also a very small $O(\alpha_s)$ normal component
of the polarization vector which will contribute to $|\vec{P}\,|$ at the
$O(0.01)$. It is so small that it is not discernible in our numerical plots.

In Fig.~\ref{f:fig16} we show a plot of the energy dependence of the polar
average $\langle\,|\vec{P}\,|\,\rangle$ of the total polarization. For both
$P_{\rm eff}=\pm1$ the average polarization is large in the whole energy
range with a slight decrease when the energy is increased. The $P_{\rm eff}=+1$
polarization is slightly larger than the $P_{\rm eff}=-1$ polarization. The
average total polarization becomes smaller when the effective polarization
is reduced from $P_{\rm eff}=\pm1$. As has been discussed before the rate of
decrease is much faster for $P_{\rm eff}=+1$ than for $P_{\rm eff}=-1$ as can
be appreciated by comparing the $P_{\rm eff}=-0.5$ and $P_{\rm eff}=+0.5$
curves. The smallest polarization in Fig.~\ref{f:fig16} is obtained for
$P_{\rm eff}=+0.5$. As will be discussed further on $P_{\rm eff}=+0.5$ is
close to the effective beam polarization where one has minimal polarization.

\begin{figure}[t]
\begin{center}
\epsfig{figure=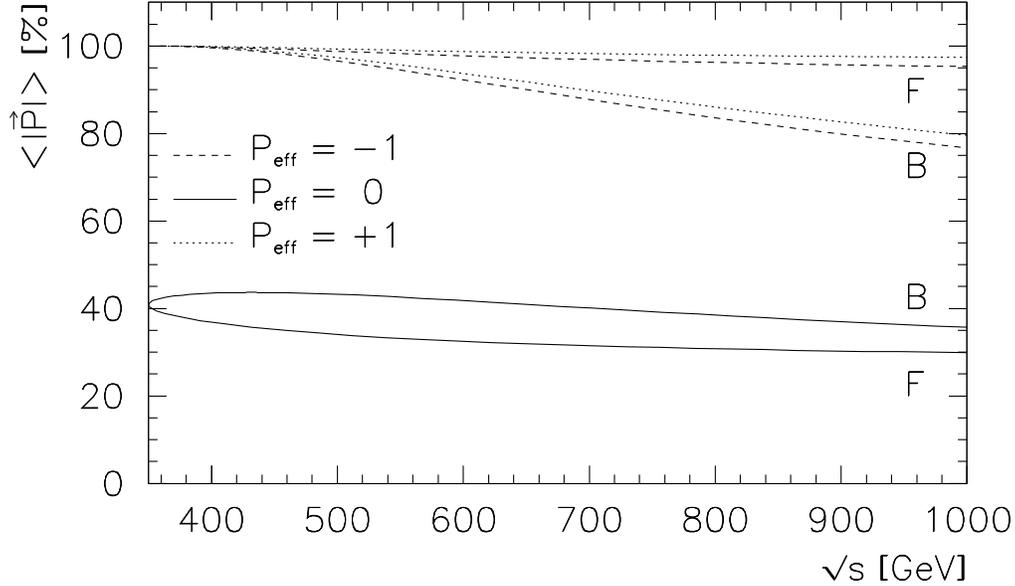, scale=0.8}
\caption{Total NLO top quark polarization averaged over the forward and 
  backward hemispheres for $P_{\rm eff}=-1,0,+1$ as a function of $\sqrt s$
  \label{polafb}}
\end{center}
\end{figure}

Of interest is the total polarization in the forward and backward hemispheres.
In Fig.~\ref{polafb} we show plots of the average total polarization
$\langle\,|\vec{P}\,|\,\rangle_{F/B}$ for $P_{\rm eff}=\pm1,0$, where the
averaging is done over the forward and backward hemispheres. The average
total polarization $\langle\,|\vec{P}\,|\,\rangle_F$ in the forward hemisphere
is quite large for both $P_{\rm eff}=\pm1$ and remains larger than $95\%$ even
up to $\sqrt s=1000\GeV$. This is quite welcome from the point of view of
statistics since the bulk of the rate is in the forward hemisphere. The
$P_{\rm eff}=+1$ polarization is slightly larger than the $P_{\rm eff}=-1$
polarization. The average backward polarization
$\langle\,|\vec{P}\,|\,\rangle_B$ is significantly smaller than the forward
polarization $\langle\,|\vec{P}\,|\,\rangle_F$ for both $P_{\rm eff}=\pm1$ as
can also be appreciated by looking at Fig.~\ref{polah}. Both forward and
backward $P_{\rm eff}=0$ polarizations show a slightly decreasing energy
behaviour starting at the common threshold value of
$\langle\,|\vec{P}\,|\,\rangle=A_{RL}=0.409$.

Returning to Fig.~\ref{polah}b ($P_{\rm eff}=+1$), one observes a
conspicuously large $10\%$ radiative correction at the backward point for
$\sqrt{s}=3000\GeV$ where the Born term prediction is $|\vec{P}\,|=1$. One
can attempt to understand this large value by substituting the asymptotic
values of the radiative corrections calculated in Ref.~\cite{Groote:2009zk}.
For the surviving longitudinal component $P^{(\ell)}$, one obtains
\begin{equation}
\label{radcorR}
P^{(\ell)}= \,-\,\bigg(\,1-\frac{\alpha_{s}}{3\pi}\,
\Big(\,\frac{f_{RR}^{2}}{f_{RL}^{2}}+[2]\,\Big)\,+\,\ldots\,\bigg)\,,
\end{equation}
where the bracketed notation ``$[2]$'' denotes the anomalous contribution not
present in $m_t/\sqrt{s}=0$ production (see Ref.~\cite{Groote:2009zk}). 
Using $f_{RR}^2/f_{RL}^2=16.069$ and $\alpha_s(3000\GeV)=0.079$ the radiative
correction at the backward point amounts to $15\%$ which is reasonably close
to the value in Fig.~\ref{polah}b. The anomalous contribution is quite small.
The corresponding formula for Fig.~\ref{polah}a ($P_{\rm eff}=-1$) is
obtained from Eq.~(\ref{radcorR}) by the substitution
$f_{RR,\,RL}\,\to\,f_{LL,\,LR}$. With $f_{LL}^2=1.417$ and $f_{LR}^2=0.188$,
one obtains a radiative correction of $8\%$ at the backward point, again in
approximate agreement with Fig.~\ref{polah}a. One may state that the large
radiative corrections at the backward point for $P_{\rm eff}=\pm1$ at
$\sqrt s=3000\GeV$ result from the fact that $f_{RR}\gg f_{RL}$ and
$f_{LL}\gg f_{LR}$.  

\begin{figure}
\begin{center}
\epsfig{figure=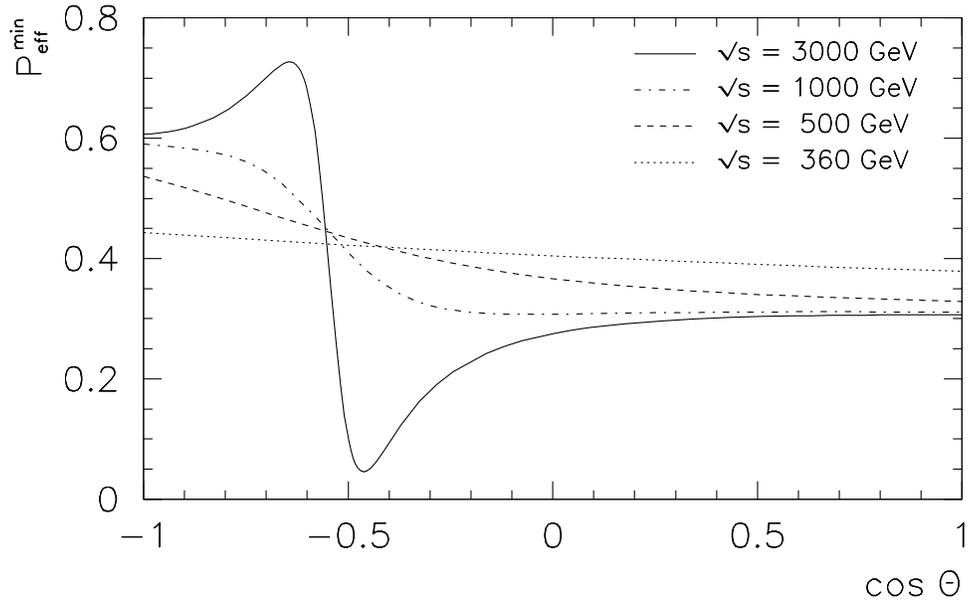, scale=0.8}\\[12pt]\qquad(a)\\[12pt]
\epsfig{figure=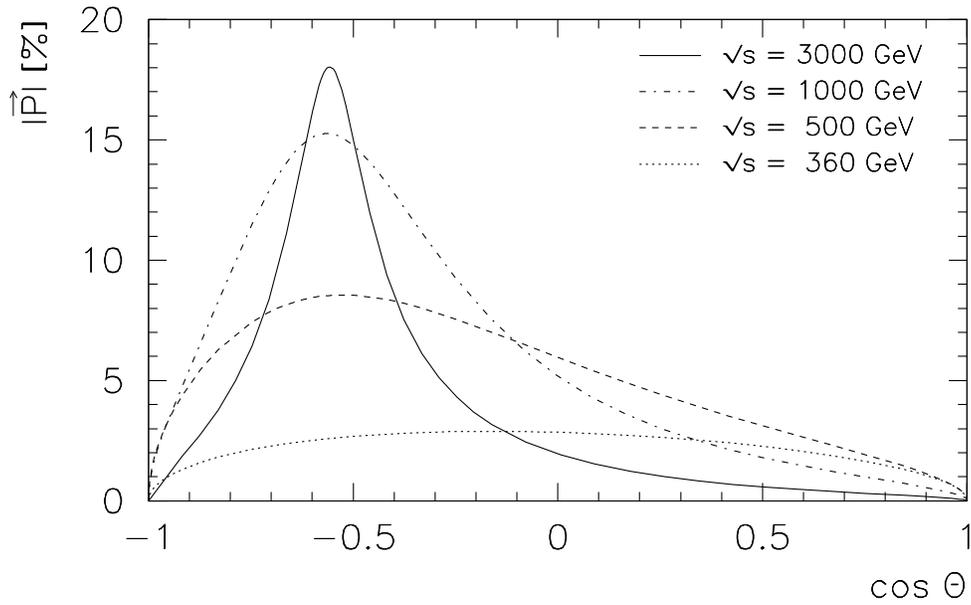, scale=0.8}\\[12pt]\qquad(b)
\end{center}
\caption{(a) $P_{\rm eff}$ values required for minimal top quark polarization
  $|\vec{P}\,|$ and (b) minimal values for $|\vec{P}\,|$, plotted against
  $\cos\theta$, for $\sqrt s=360$, $500$, $1000$, and $3000\GeV$ (notation as
  in Fig.~\ref{sigmar})
  \label{f:fig13}}
\end{figure}

Next we investigate the parameter space for which the polarization of the top
quark is minimal. For some measurements it may be advantageous to eliminate or
minimize polarization effects. For once, one can thereby gauge the efficiency
of a polarization measurement against an unpolarized control sample. The
parameters to be varied are the effective beam polarization $P_{\rm eff}$, the
polar angle $\theta$, and the energy $\sqrt s$. The minimization is done at NLO
including the normal polarization component according to Eq.~(\ref{totpol}). 

In Fig.~\ref{f:fig13}a we show a plot of the NLO values of $P_{\rm eff}$ which
minimize $|\vec{P}\,|$ for any given scattering angle. The minimizing values
$P_{\rm eff}^{{\rm min}}$ depend in addition on the energy. An important
feature of the minimizing effective beam polarization is that, in the forward
region, where the rate is largest, the dependence of $P_{\rm eff}^{\rm min}$
on $\cos\theta$ is reasonably flat for all shown energies. This means that it
is possible to tune the effective beam polarization in the forward region for
each energy such that one obtains approximate minimal polarization. Just above
threshold at $\sqrt s=360\GeV$, $P_{\rm eff}^{{\rm min}}$ is close to the flat
behaviour at threshold $P_{\rm eff}^{\rm min}=A_{RL}=0.409$. Apart from the
near-threshold curve $P_{\rm eff}^{\rm min}$ shows a strong dependence on
$\cos\theta$ in the backward region. The corresponding minimal values of
$|\vec{P}\,|$ are shown in Fig.~\ref{f:fig13}b. At the forward and backward
point the minimal polarization is zero by construction. In the forward region
the polarization remains quite small starting from zero at the forward point.
This is different in the backward region where the minimal polarization can
become as large as $18\%$ for the highest shown energy of $\sqrt s=3000\GeV$.

\begin{figure}
\begin{center}
\epsfig{figure=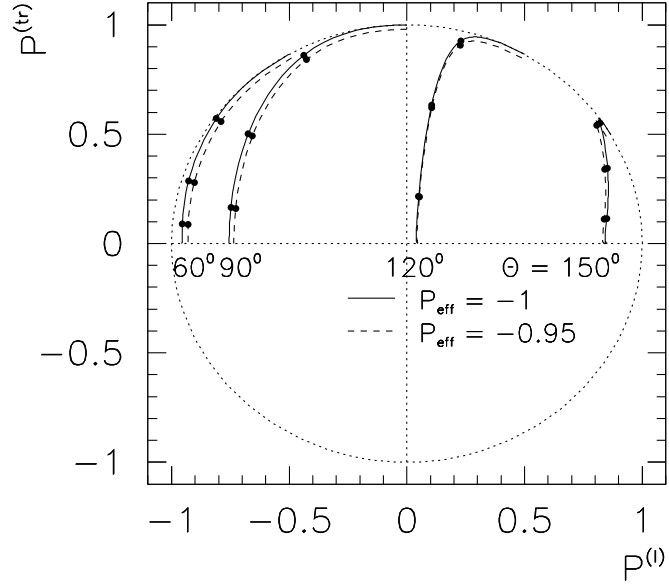, scale=0.8}\\[12pt]\qquad(a)\\[12pt]
\epsfig{figure=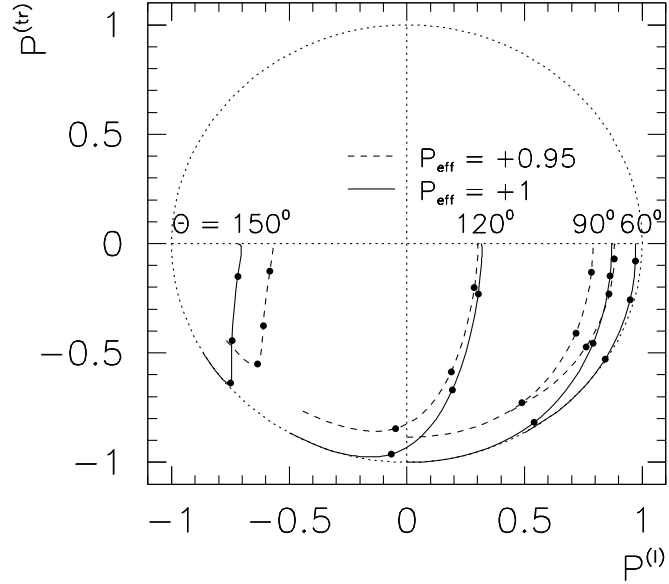, scale=0.8}\\[12pt]\qquad(b)
\end{center}
\caption{NLO Parametric plot of the orientation and the length of the
  polarization vector in dependence on the c.m.\ energy $\sqrt s$ for values
  $\theta=60^\circ$, $90^\circ$, $120^\circ$, and $150^\circ$ for (a)
  $P_{\rm eff}=-1$ (solid lines) and $P_{\rm eff}=-0.95$ (dashed lines), and
  (b) $P_{\rm eff}=+1$ (solid lines) and $P_{\rm eff}=+0.95$ (dashed
  lines). The dots on the trajectories from the border to the central line
  stand for $\sqrt{s}=500$, $1000$, and $3000\GeV$
  \label{pardeps}}
\end{figure}

We now turn to the orientation of the polarization vector. We have already
discussed some aspects of the orientation of the polarization vector of the
top quark in Secs.~4 and 5. We now combine the information on the orientation
and the magnitude of the polarization vector in one single 
(radius, angle) plot where we trace the end point (apex) of the polarization 
vector $\vec{P}$ within the unit circle while increasing the energy from
threshold to infinity. The apex stays within the unit circle since
$|\vec{P}\,|\le 1$. In Fig.~\ref{pardeps}a we consider the case
$P_{\rm eff}=-1$ for the polar angles $\theta=60^\circ$, $90^\circ$,
$120^\circ$, and $150^\circ$. All trajectories start off at threshold where
$|\vec{P}\,|=1$ and $\alpha=180^\circ-\theta$ and, depending on $\cos\theta$,
end up at $\alpha=0^\circ$ or $\alpha=180^\circ$ with a length close to the
asymptotic Born term result (Eq.~(\ref{Lhelimit})). Which of the two asymptotic
solutions $\alpha=0^\circ$ and $\alpha=180^\circ$ are attained can be traced
to the corresponding LO result (Eq.~(\ref{Lhelimit})) or from
Fig.~\ref{f:fig7}b. For $P_{\rm eff}=+1$ (Fig.~\ref{pardeps}b) the
trajectories start off at threshold with $\alpha=-\theta$ and end up at
$\alpha=0^\circ$ or $\alpha=180^\circ$. The appropriate solution can again be
read off from the Born term formula (Eq.~(\ref{Lhelimit})) with the
appropriate replacements as described after Eq.~(\ref{Lhelimit}), or from
Fig.~\ref{f:fig7}b. The length of the asymptotic polarization vector is close
to what is obtained from Eq.~(\ref{Lhelimit}) after the appropriate
replacements. Since $f_{LR}^2/f_{LL}^2>f_{RL}^2/f_{RR}^2$ the asymptotic
values of $|\vec{P}\,|$ and thereby the intermediate values of $|\vec{P}\,|$
are larger for $P_{\rm eff}=+1$ than for $P_{\rm eff}=-1$. We remind the
reader, though, that extrapolations away from $P_{\rm eff}=-1$ are more stable
than extrapolations away from $P_{\rm eff}=+1$. This is illustrated in
Figs.~\ref{pardeps}a and~\ref{pardeps}b by adding the corresponding
trajectories (dashed lines) for $P_{\rm eff}=-0.95$ and $P_{\rm eff}=+0.95$,
respectively. One observes only a minor change in Fig.~\ref{pardeps}a going
from $P_{\rm eff}=-1$ to $P_{\rm eff}=-0.95$. For example, the total
polarization $|\vec{P}\,|$ remains close to maximal at $\sqrt s=500\GeV$ for
the technically feasible effective beam polarization of $P_{\rm eff}=-0.95$.
The corresponding changes in Fig.~\ref{pardeps}b are much larger. In
particular, the total polarization $|\vec{P}\,|$ at $P_{\rm eff}=+0.95$ is
considerably reduced from its values at $P_{\rm eff}=+1$.

In Figs.~\ref{pardeps}a and~\ref{pardeps}b we have marked the energy
dependence of the polarization vector by dots (or ticks) on the trajectory
of the apex of the polarization vector. One notes that there is very little
change in the length of the polarization vector going from threshold to
$\sqrt{s}=500\GeV$. The ticks are approximately equally spaced on the
trajectories indicating an approximate inverse power law dependence of the
spacing on the energy. For the three trajectories  $\theta=60^\circ$,
$90^\circ$ and $120^{\circ}$, the angle $\alpha$ is monotonically increasing
with energy. In contrast to this the $\theta=150^\circ$ trajectory shows a
kink at around $\sqrt{s}=500\GeV$. Both Figs.~\ref{pardeps}a and~\ref{pardeps}b
show that at $\sqrt{s}=3000\GeV$ one has not yet reached the asymptotic regime.

\section{Longitudinal spin--spin correlations}

Up to this point we have only discussed the single-spin polarization of the
top quark. The polarizations of pair produced top and antitop quarks are
correlated and could be observed in the energy spectra of decay products,
especially in the energy spectra of leptons and antileptons. There are
altogether nine double-density matrix elements describing the spin--spin
correlations of the top and antitop quarks. Here we concentrate on the
longitudinal spin--spin correlation which is the double-density matrix element
that survives in the high-energy limit (for analytical NLO results see
Refs.~\cite{Groote:2009zk,tbp98,gkl98}). We mention that the full set of NLO
double-density matrix elements has been numerically evaluated in
Refs.~\cite{bfu99a,bfu99b}.

The longitudinal spin--spin correlation cross section is defined by
\begin{equation}
\sigma_\alpha^{(\ell_1\ell_2)}=\sigma_\alpha(\uparrow\uparrow)
  -\sigma_\alpha(\uparrow\downarrow)-\sigma_\alpha(\downarrow\uparrow)
  +\sigma_\alpha(\downarrow\downarrow)\,,
\end{equation}
where e.g.\ $(\uparrow\uparrow)$ denotes a top quark with helicity $1/2$ and 
an antitop quark with helicity $1/2$, etc. Similar to Eq.~(\ref{diffcross}),
the differential $\cos\theta$ distribution is given by
\begin{equation}
\label{diffcross1}
\frac{d\sigma^{(\ell_1\ell_2)}}{d\cos\theta}
  =\frac38(1+\cos^2\theta)\,\sigma_U^{(\ell_1\ell_2)}
  +\frac34\sin^2\theta\,\sigma_L^{(\ell_1\ell_2)} 
  +\frac34\cos\theta\,\sigma_F^{(\ell_1\ell_2)}\,,
\end{equation}
where
\begin{eqnarray}
\sigma_{U,L}^{(\ell_1\ell_2)}&=&(1-h_-h_+)\frac{\pi\alpha^2v}{3q^4}
  \left((g_{11}+P_{\rm eff}g_{41})\,H_{U,L}^{1(\ell_1\ell_2)}
  +\,(g_{12}+P_{\rm eff}g_{42})\,H_{U,L}^{2(\ell_1\ell_2)}\right)\,,
  \nonumber\\
\sigma_{F}^{(\ell_1\ell_2)}&=&(1-h_-h_+)\frac{\pi\alpha^2 v}{3q^4}
  \Big(g_{44}+P_{\rm eff}g_{14}\Big)\,H_{F}^{4(\ell_1\ell_2)}\,.
\end{eqnarray}
The Born term contributions read~\cite{Groote:2009zk}
\begin{eqnarray}
H_U^{1(\ell_1\ell_2)}({\it Born\/})=-2N_cq^2(1+v^2),&&
H_L^{1(\ell_1\ell_2)}({\it Born\/})=N_cq^2(1-v^2)
  =H_L^{2(\ell_1\ell_2)}({\it Born\/}),\nonumber\\[3pt]
H_U^{2(\ell_1\ell_2)}({\it Born\/})=-2N_cq^2(1-v^2),&&
H_F^{4(\ell_1\ell_2)}({\it Born\/})=-4N_cq^2v.
\end{eqnarray}
Note that one has the Born term relations
\begin{eqnarray}
\label{born1}
H_U^{1,2}({\it Born})&=&-H_U^{1,2\,(\ell_1\ell_2)}({\it Born}),\nonumber\\
H_F^4({\it Born})&=&-H_F^{4\,(\ell_1\ell_2)}({\it Born}),\nonumber\\
H_L^{1,2}({\it Born})&=&H_L^{1,2\,(\ell_1\ell_2)}({\it Born})\,,
\end{eqnarray}
which are due to angular momentum conservation in the back-to-back
configuration of the Born term production~\cite{Groote:2009zk}. These
relations no longer hold true in the case of additional gluon emission. The
relations~(\ref{born1}) imply that $P^{(\ell_1\ell_2)}=-1$ at
$\cos\theta=\pm1$ independent of $P_{\rm eff}$. Since the transverse
contributions $H_{U,F}$ dominate over the longitudinal contributions $H_L$ one
anticipates from the relations~(\ref{born1}) that the longitudinal
spin--spin correlations are negative and only weakly beam polarization
dependent.

Similar to Eq.~(\ref{NoverD}), the $\cos{\theta}$ dependent longitudinal
spin--spin correlation is defined by the ratio
\begin{equation}
P^{(\ell_1\ell_2)}(\cos\theta)
  =\frac{N^{(\ell_1\ell_2)}(\cos\theta)}{D(\cos\theta)}\,,
\end{equation}
with the denominator function given in Eq.~(\ref{D}). The numerator function
is given by
\begin{eqnarray}
N^{(\ell_1\ell_2)}(\cos\theta) &=&\frac38(1+\cos^2\theta)
  \left((g_{11}+g_{41}P_{\rm eff})H_U^{1(\ell_1\ell_2)}
  +(g_{12}+g_{42}P_{\rm eff})H_U^{2(\ell_1\ell_2)}\,\right)\nonumber\\&&
  +\frac34\sin^2\theta\left((g_{11}+g_{41}P_{\rm eff})H_L^{1(\ell_1\ell_2)}
  +(g_{12}+g_{42}P_{\rm eff})H_L^{2(\ell_1\ell_2)}\,\right)\nonumber\\&&
  +\frac34\cos\theta\,\,(g_{44}+g_{14}P_{\rm eff})H_F^{4(\ell_1\ell_2)}\,.
\end{eqnarray}

Let us first consider the polar angle average of the longitudinal spin--spin
correlation. For the Born term contribution, one obtains
\begin{equation}
\langle P^{(\ell_1\ell_2)}\rangle 
  =-\frac13\,\,\frac{(g_{11}+P_{\rm eff}\,g_{41})(1+3v^2)
  +(g_{12}+P_{\rm eff}\,g_{42})(1-v^2)}{(g_{11}+P_{\rm eff}\,g_{41})(1+v^2/3)
  +(g_{12}+P_{\rm eff}\,g_{42})(1-v^2)}\,.
\end{equation}
Note that $\langle P^{(\ell_1\ell_2)}\rangle=-1/3$ at threshold ($v=0$) and
$\langle P^{(\ell_1\ell_2)}\rangle=-1$ in the high-energy limit ($v=1$)
independent of the beam polarization parameter $P_{\rm eff}$. In fact, the
dependence on $P_{\rm eff}$ is very weak also for energies intermediate
between these two limits. For example, for $\sqrt s=500\GeV$ one finds
\begin{equation}
\label{ellell1}
\langle P^{(\ell_1\ell_2)}\rangle
  =\langle P^{(\ell_1\ell_2)}\rangle(P_{\rm eff}=0)\,\,
  \frac{1-0.36P_{\rm eff}}{1-0.37P_{\rm eff}}
  =\left\{\begin{array}{rl}
  -0.67&\quad P_{\rm eff}=+1\\
  -0.65&\quad\phantom{P_{\rm eff}}=\phantom{+}0\\
  -0.65&\quad\phantom{P_{\rm eff}}=-1
  \end{array}\right\}\,.
\end{equation}
Equation~(\ref{ellell1}) shows that the dependence on the beam polarization
parameter $P_{\rm eff}$ practically drops out in the ratio~(\ref{ellell1}).

\begin{figure}[t]
\begin{center}
\epsfig{figure=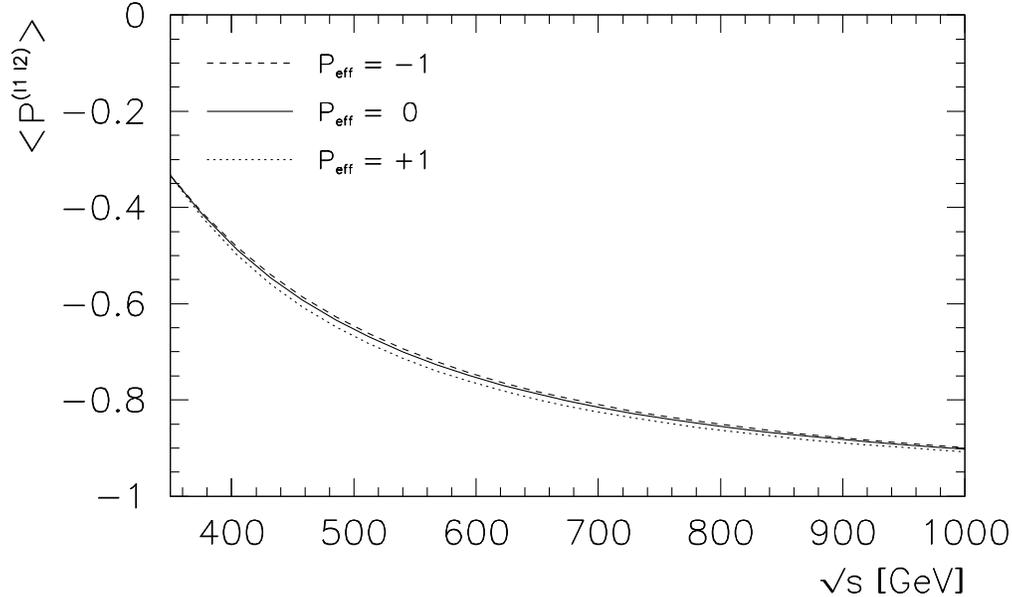, scale=0.8}
\end{center}
\caption{NLO beam energy dependence of the polar average of the longitudinal 
  spin--spin correlation function $\langle P^{(\ell_1\ell_2)}\rangle$
  \label{f:fig20}}
\end{figure}

In Fig.~\ref{f:fig20} we plot the average longitudinal spin--spin correlation
function $\langle P^{(\ell_1\ell_2)}\rangle$ up to $O(\alpha_s)$ as a function
of $\sqrt s$ for different effective beam polarizations. The dependence of
$\langle P^{(\ell_1\ell_2)}\rangle$ on $P_{\rm eff}$ is barely visible.

As shown in Ref.~\cite{ps96} the off-diagonal basis defined by
Eq.~(\ref{offdiag}) diagonalizes both the single-spin and the spin--spin
double-density matrix at the Born term level for $P_{\rm eff}=-1$. In the NLO
calculation described in this section the top and antitop quark are no longer
back to back due to hard gluon emission, i.e.\ in the two helicity basis (top
and antitop quark) the two respective $z$ axis are not in general
back to back. In the high-energy limit, where only the longitudinal spin--spin
density matrix elements survive, the NLO spin--spin density matrix elements
are therefore not simply related to the off-diagonal basis introduced in
Ref.~\cite{ps96}. A discussion of the rigidity of back-to-back
$(t\bar t)$ pairs with respect to gluon emission in $e^+e^-$ collisions can be
found in Ref.~\cite{Groote:1998xc}.

\section{Summary and conclusions}

We have discussed in detail top quark polarization in above-threshold
$(t\bar t)$ production at a polarized linear $e^+e^-$ collider within the SM.
While journeying through the three-dimensional
$(P_{\rm eff},\,\cos\theta,\,\sqrt s)$-parameter space a rich landscape of SM
polarization phenomena unfolds which awaits experimental confirmation or
falsification. Generally speaking, one needs large values of the effective
beam polarization if the aim is to produce highly polarized top quarks. Very
small or zero polarization of the top quark can be obtained by fine-tuning the
parameters $(P_{\rm eff},\,v,\,\cos\theta)$.  

The $(t\bar t)$-production rate at a polarized linear $e^+e^-$ collider is
governed by the gain factor $K_G=1-h_-h_+$ and the effective beam polarization
$P_{\rm eff}$. The optimal choice as concerns the rate is $h_-$ negative and
$h_+$ positive such that one has $K_G>1$ and $P_{\rm eff}<0$, i.e.\ the
optimal choice for the rate would lie in the second quadrant of the
$(h_-,h_+)$ plane in Fig.~\ref{f:fig3}. The largest gain in the rate is
obtained for $h_-=-h_+=-1$, i.e.\ for $K_G=2$ and $P_{\rm eff}=-1$.

At (Born term) threshold, one has a flat $\cos\theta$ distribution. As the
energy increases there is a quick turn into forward dominance, with little
dependence on $P_{\rm eff}$. This is a welcome feature for polarization
measurements, which require large statistics and rates, since forward
production is advantageous for stable and large top quark spin effects. More
explicitly, the polarization of the top quark is generally large and more
stable against variations of the parameters $P_{\rm eff}$, $\cos\theta$ and
the energy in the forward region than in the backward region.

Contrary to the rate, the polarization observables depend only on
$P_{\rm eff}$, and not separately on $h_-$ and $h_{+}$. We find that the
single-spin polarization of the top quark is, in general, strongly dependent
on the effective beam polarization parameter $P_{\rm eff}$. This is quite
different for longitudinal spin--spin correlations which depend only weakly on
beam polarization effects.

In order to attain small or large values of the polarization would in general
require an extreme fine-tuning of $P_{\rm eff}$ depending on $\cos\theta$ and
the energy. The good news is that the polarization properties at
$\sqrt s=500\GeV$ are still quite close to the polarization properties at
threshold where they are quite simple. If the aim is to achieve zero or small
polarization at $\sqrt s=500\GeV$ a choice of $P_{\rm eff}=0.36\div0.40$ leads
to very small values of $|\vec{P}\,|$ in the forward hemisphere where the
rate is largest. At $\sqrt s=500\GeV$, close to maximal values of the
polarization $|\vec{P}\,|\simeq 1$ can be achieved over the whole
$\cos\theta$ range for effective beam polarizations close to $P_{\rm eff}=-1$
or $P_{\rm eff}=+1$, where the polarization is slightly larger for
$P_{\rm eff}=+1$. However, a choice close to $P_{\rm eff}=-1$ is preferred
because of two reasons. First, this choice leads to larger rates and, second,
the polarization observables are more stable against variations of
$P_{\rm eff}$ close to $P_{\rm eff}=-1$ than close to $P_{\rm eff}=+1$. For
$\sqrt s=1000\GeV$ a total polarization of $|\vec{P}\,|>85\%$ and
$|\vec{P}\,|>90\%$ can be achieved in the forward hemisphere for
$P_{\rm eff}=-1$ and $P_{\rm eff}=+1$, respectively. The highest energy
considered in this paper is $\sqrt s=3000\GeV$. We have found that the
polarization results at $\sqrt s=3000\GeV$ are, in many aspects, not very
close to their respective asymptotic values.

For the analysis of polarization effects one also needs to know the orientation
of the polarization vector. We have given explicit results on its orientation
where we have found that, at $\sqrt s=500\GeV$, the polarization vector is
still approximately aligned or counteraligned with the electron momentum as is
the case at threshold.

Our results can be viewed as a generalization of the $P_{\rm eff}= -1$ results
of Ref.~\cite{ps96} to general values of $-1\le P_{\rm eff}\le+1$. We have
checked that all our Born term formulas agree with those of Ref.~\cite{ps96}
when we set $P_{\rm eff}= -1$ in our Born term expressions. In addition, we
have derived simple Born term rate and polarization formulas for the case
$P_{\rm eff}=+1$ not treated explicitly in Ref.~\cite{ps96}. We also provide
$O(\alpha_s)$ corrections to the Born term results which we have checked
against the corresponding $O(\alpha_s)$ corrections in the helicity system
given in Ref.~\cite{Kodaira:1998gt}. In addition, we provide radiative
corrections to the orientation angle $\alpha$ of the polarization vector which
were not discussed in Ref.~\cite{Kodaira:1998gt}.

All the results in this paper refer to the polarization of the top quark. In
order to obtain the SM coupling predictions for the polarization of the
antitop quark, let us first set up an orthonormal spin basis for the antitop
quark by replacing the momenta in Eq.~(\ref{basistop}) by their charge
conjugate partners, i.e.\ $\vec p_t\to\vec p_{\bar t}$ and
$\vec p_{e^-}\to\vec p_{e^+}$. The three orthonormal basis vectors
$(\vec e^{\,(tr)},\vec e^{\,(n)},\vec e^{\,(\ell)})$ are now given by
\begin{equation}
\label{basisantitop}
\vec e^{\,(tr)}=\frac{(\vec p_{e^+}\times\vec p_{\bar t})\times\vec p_{\bar t}}
{|(\vec p_{e^+}\times\vec p_{\bar t})\times\vec p_{\bar t}|},\qquad
\vec e^{\,(n)}=\frac{\vec p_{e^+}\times\vec p_{\bar t}}
{|\vec p_{e^+}\times\vec p_{\bar t}|},\qquad 
\vec e^{\,(\ell)}=\frac{\vec p_{\bar t}}{|\vec p_{\bar t}|}.
\end{equation}
In the polar angle distribution~(\ref{diffcross}) the polar angle now refers
to $\theta_{\bar te^-}$ and {\it not} to $\theta=\theta_{te^-}$, as in the top
quark case discussed in the main part of this paper. Since the lepton pair is
back to back in the lab frame, one has
$\theta_{\bar te^-}=180^\circ-\theta_{\bar te^+}$, i.e.\ the two terms in
Eq.~(\ref{diffcross}) proportional to $\cos\theta$ change sign if written in
terms of $\cos\theta_{\bar t e^+}$. In the SM the rate and the polarization
components of the antitop quark are related to those of the top quark via
\begin{eqnarray}
\sigma_{\bar t}(\cos\theta_{\bar te^+})&=&\sigma_t(\cos\theta_{te^-})\,,
\nonumber \\
P_{\bar t}^{(\ell,n)}(\cos\theta_{\bar te^+})
  &=&-P_t^{(\ell,n)}(\cos\theta_{te^-}) \,,\nonumber \\ 
P_{\bar t}^{(tr)}(\cos\theta_{\bar te^+})
  &=&P_t^{(tr)}(\cos\theta_{te^-})\,.
\end{eqnarray}
As an example, and as expected, the antitop quark is predominantly produced
in the backward hemisphere relative to the $e^-$ direction.

In polarized top decay the compositions of helicity fractions of the final
state $W^-$ bosons change relative to the helicity fractions of unpolarized
top quark decay depending on the magnitude and orientation of the polarization
vector. This polarization effect has been investigated in a number of papers
where a variety of spin observables have been defined which involve the
dominant decay mode of the top quark $t(\uparrow)\to b+W^+(\to \ell^++\nu_l)$.
The analysis can be done in the $(e^+ e^-)$ c.m.\ frame as in
Ref.~\cite{Grzadkowski:2000nx}, in the top quark rest frame as e.g.\ in
Refs.~\cite{Czarnecki:1993gt,Groote:2006kq,Korner:1998nc}, or in the $W$ rest
frame as e.g.\ in
Refs.~\cite{Fischer:1998gsa,Fischer:2001gp,AguilarSaavedra:2010nx}.
References~\cite{Fischer:1998gsa,Czarnecki:1993gt,Korner:1998nc,Fischer:2001gp}
concentrate on SM predictions and discuss radiative
QCD~\cite{Fischer:1998gsa,Czarnecki:1993gt,Fischer:2001gp} corrections
to the respective spin observables, while Ref.~\cite{Grzadkowski:2000nx}
analyzes the effect of non-SM interactions in the production and decay of the
top quark. The authors of Ref.~\cite{AguilarSaavedra:2010nx} discuss some
novel spin observables and proceed to analyze the effect of non-SM decay
vertices on these observables. QCD corrections to non-SM interactions in the
decay of an unpolarized top quark have been recently calculated in
Ref.~\cite{Drobnak:2010ej}. This calculation can be easily extended to
polarized top quark decay.

The discussion of this paper has focused on SM physics with longitudinal beam
polarization. Non-SM electroweak couplings on the production side, involving
leptons and quarks, and transverse beam polarization effects can be easily
included using the formalism of this paper. Transverse beam polarization
effects will be discussed in a sequel to this paper.

\subsection*{Acknowledgments}
S.G.\ acknowledges the support by the Estonian target financed Project
No.~0180056s09 and by the Deutsche Forschungsgemeinschaft (DFG) under Grant
No.~436~EST~17/1/06. B.M.\ acknowledges the support of the Ministry of Science
and Technology of the Republic of Croatia under Contract No.~098-0982930-2864.
S.P.\ is supported by the Slovenian Research Agency and by the European RTN
network FLAVIAnet (Contract No.~MRTN-CT-035482).

\section*{Appendix: SM values of the\\ electroweak coupling coefficients}
\setcounter{equation}{0}\def\theequation{A\arabic{equation}}
The electroweak coupling matrix elements $g_{ij}(s)$ needed in this paper are
given by
\begin{eqnarray}
\label{ewcoefficients}
g_{11/12}&=&Q_f^2-2Q_fv_ev_f\real\cz+(v_e^2+a_e^2)
(v_f^2\pm a_f^2)|\cz|^2 \quad(\,=0.61/0.34)\,,\nonumber\\
g_{14}&=&2Q_fv_ea_f\real\cz-2(v_e^2+a_e^2)v_fa_f|\cz|^2 \quad(\,=-0.14),
\nonumber\\
g_{41/42}&=&2Q_fa_ev_f\real\cz-2v_ea_e(v_f^2\pm a_f^2)|\cz|^2 \quad
(\,=-0.21/\!\!-0.17),\nonumber\\
g_{44}&=&-2Q_fa_ea_f\real\cz+4v_ea_ev_fa_f|\cz|^2 \quad(\,=0.50)\,,
\nonumber
\end{eqnarray}
where 
\begin{equation}
\cz(s)=\frac{gM_Z^2s}{(s-M_Z^2+iM_Z\Gamma_Z)}\,,
\end{equation}
with $M_Z$ and $\Gamma_Z$ the mass and width of the $Z^0$ and
$g=(16\sin^2\theta_{W}\cos^2\theta_WM_Z^2)^{-1}
= 4.229\cdot 10^{-5}\GeV^{-2}$ where we have used $\sin^2\theta_W=0.23116$.
$Q_f$ are the charges of the final state quarks to which the electroweak
currents directly couple; $v_e$ and $a_e$, $v_f$ and $a_f$ are the electroweak
vector and axial vector coupling constants. For example, in the Weinberg-Salam
model, one has $v_e=-1+4\sin^2\theta_W$, $a_e=-1$ for leptons,
$v_f=1-\frac83\sin^2\theta_W$, $a_f=+1$ for up-type quarks ($Q_f=+\frac23$),
and $v_f=-1+\frac43\sin^2\theta_W$, $a_f=-1$ for down-type quarks
($Q_f=-\frac13$). The electroweak coupling coefficients $g_{ij}$ are not
independent. They satisfy the constraints
\begin{equation} 
\Big((g_{11}\pm g_{41})^2-(g_{14}\pm g_{44})^2\Big)^{1/2}=g_{12}\pm g_{42}\,.
\end{equation}

In Eq.~(\ref{ewcoefficients}) we have also listed the numerical values of the
electroweak coefficients for $(t\bar t)$ production at $\sqrt s=500\GeV$. As
already mentioned in the main text, it is safe to work in the zero width
approximation for the $Z$ boson above $(t \bar t)$ threshold, i.e.\ we set
$\Gamma_Z=0$. Note that the numerical values of the electroweak coefficients
are only weakly energy dependent above the $(t \bar t)$ threshold. The energy
dependence comes from the energy-dependent factor $\cz(s)$ which takes the
values $0.377$, $0.364$, and $0.352$ for $\sqrt s=350\GeV$ (threshold),
$500\GeV$, and infinite energy, respectively.

For some applications it is convenient to switch to chiral representations
of the initial and final electromagnetic and weak currents as was done in
Ref.~\cite{ps96}. Accordingly, one defines coefficients
\begin{equation}
f_{LL/LR}=-Q_f+(v_e+a_e)(v_f\pm a_f)\cz(s)\,.
\end{equation}
The chiral electroweak coefficients $f_{LL/LR}$ can be seen to be related to
the above $g_{ij}$ via
\begin{eqnarray}
f_{LL/LR}&=&-(g_{11}\mp g_{14}-g_{41}\pm g_{44})^{1/2}\quad
  (\,=-1.21/\!\!-0.43)\,,
\nonumber\\
f_{LL}f_{LR}&=&g_{12}-g_{42}\qquad\qquad\qquad\qquad\qquad(\,=0.51)\,.
\end{eqnarray}
For the case $P_{\rm eff}=+1$, one also needs the corresponding relations
for the coefficients
\begin{equation}
f_{RR/RL}=-Q_f+(v_e-a_e)(v_f\mp a_f)\cz(s)\,.
\end{equation}
One has
\begin{eqnarray}
f_{RR/RL}&=&-(g_{11}\pm g_{14}+g_{41}\pm g_{44})^{1/2}\quad
  (\,=-0.87/\!\!-0.20)\,,
\nonumber\\
f_{RR}f_{RL}&=&g_{12}+g_{42}\qquad\qquad\qquad\qquad\qquad(\,=0.18)\,.
\end{eqnarray}

\end{document}